\documentclass[aps,pra,reprint]{revtex4-2}
\usepackage[normalem]{ulem}
\usepackage{xcolor}
\usepackage{hyperref}

\usepackage{graphicx}
\usepackage{amsmath,amssymb,amsfonts}
\usepackage{mathrsfs} 
\usepackage{physics}  

\renewcommand{\vec}[1]{\mathbf{#1}}
\newcommand{\rb}{\mathbf{r}}

\newcommand{\Power}{\mathscr{P}}

\newcommand{\dr}{\mathbf{r}-\mathbf{r}'}

\newcommand{\je}{\mathbf{j}_{e}}
\newcommand{\jm}{\mathbf{j}_{m}}

\newcommand{\Te}{\boldsymbol{\mathcal{T}}_{-}}
\newcommand{\Ti}{\boldsymbol{\mathcal{T}}_{+}}
\newcommand{\Ke}{\boldsymbol{\mathcal{K}}_{-}}
\newcommand{\Ki}{\boldsymbol{\mathcal{K}}_{+}}

\newcommand{\Tie}{\boldsymbol{\mathcal{T}}_{\pm}}
\newcommand{\Kie}{\boldsymbol{\mathcal{K}}_{\pm}}

\newcommand{\eo}{\mathbf{e}_{0}}
\newcommand{\ho}{\mathbf{h}_{0}}

\newcommand{\Zi}{\zeta^{+}}
\newcommand{\Ze}{\zeta^{-}}

\begin{document}

\title{Modified Langevin noise formalism for multiple quantum emitters \\ in dispersive electromagnetic environments out of equilibrium}

\author{Giovanni Miano}
\affiliation{Scuola Superiore Meridionale, via Mezzocannone 4,  Napoli, 80125, Italy}
\affiliation{Department of Electrical Engineering and Information Technology, Universit\`{a} degli Studi di Napoli Federico II, via Claudio 21,  Napoli, 80125, Italy}
\author{Loris Maria Cangemi}
\affiliation{Department of Electrical Engineering and Information Technology, Universit\`{a} degli Studi di Napoli Federico II, via Claudio 21,  Napoli, 80125, Italy}
\author{Carlo Forestiere}
\email[]{carlo.forestiere@unina.it}
\affiliation{Department of Electrical Engineering and Information Technology, Universit\`{a} degli Studi di Napoli Federico II, via Claudio 21,  Napoli, 80125, Italy}

\begin{abstract}
The control of interactions among quantum emitters through nanophotonic structures offers significant opportunities for quantum technologies. However, a rigorous theoretical description of the interaction of multiple quantum emitters with complex, dispersive dielectric objects remains challenging. Here, we introduce an approach based on the modified Langevin noise formalism that unveils the roles of both the noise polarization currents of the dielectrics and the vacuum fluctuations of the electromagnetic field scattered by the dielectrics. This work extends Refs. \cite{miano_quantum_2025} and \cite{miano_spectral_2025} to the general case of an arbitrary number of emitters. The proposed approach allows us to describe the dynamics of the quantum emitters for arbitrary initial quantum states of the electromagnetic environment, consisting of two independent bosonic reservoirs, a medium-assisted reservoir and a scattering-assisted reservoir, each characterized by its own spectral density matrix. Specifically, we examine situations where both reservoirs are initially in thermal quantum states but have different temperatures. Understanding how these reservoirs shape the dynamics of the emitters is crucial for understanding light–matter interactions in complex electromagnetic environments and for improving intrinsic emitter properties within structured environments. 
\end{abstract}

\maketitle

\section{Introduction}

The design of qubit-qubit interactions is fundamental to a wide range of quantum technologies, including quantum networking, quantum information processing, and quantum computation (e.g., \cite{reiserer_cavity-based_2015}, \cite{nguyen_quantum_2019}, \cite{sheremet_waveguide_2023}, \cite{qin_quantum_2024}).
Nanophotonics enables the engineering and control of quantum properties of light by embedding quantum emitters within nanostructured environments. In particular, systems that utilize quantum emitters as qubits have attracted growing interest \cite{gonzalez-tudela_lightmatter_2024, sanchez-barquilla_few-mode_2022}, with significant efforts devoted to tailoring their mutual interactions through metallic or dielectric nanostructures \cite{miguel-torcal_multiqubit_2024}. These systems are physically rich: metal and dielectric nanoparticles, typically dispersive and lossy, can support plasmonic and dielectric resonances of various multipolar orders, which can be engineered by adjusting the shape and spatial arrangement of the nanoparticles, or by exploiting collective resonances. Moreover, nanophotonic devices enable subwavelength confinement of light, allowing device miniaturization, enhanced light–matter interactions, and consequently faster dynamics and higher operating speeds. Such engineered couplings play a central role in the generation and manipulation of non-classical states of light. Nevertheless, providing a theoretical description of multiple quantum emitters that interact with realistic nanostructures remains highly nontrivial. The electromagnetic environment is inherently dispersive, and its modes span a high-dimensional continuum. Accounting for the finite extent, material losses, and dispersion of the dielectric components poses serious challenges to the canonical quantization of the electromagnetic field. In this context, macroscopic quantum electrodynamics (e.g. \cite{gruner_green-function_1996}, \cite{scheel_macroscopic_2008}), offers a widely adopted phenomenological framework, enabling a consistent and flexible quantization scheme in the presence of complex dielectric media.

Macroscopic quantum electrodynamics relies on the \textit{Langevin noise formalism}, which is based on the fluctuation-dissipation theorem. The electromagnetic field, commonly referred to as the \textit{medium-assisted field}, emerges from the dielectric noise polarization current, mediated by the dyadic Green function of the dielectric objects. As argued in \cite{stefano_mode_2001} and \cite{drezet_quantizing_2017}, the original macroscopic quantum electrodynamic model disregards the influence of vacuum fluctuations of the electromagnetic field scattered by the dielectric objects, called the \textit{scattering-assisted field} in \cite{na_numerical_2023} and \cite{ciattoni_quantum_2024}. The \textit{modified Langevin noise formalism} adds the scattering-assisted field to the medium-assisted field: polarization current fluctuations and vacuum fluctuations of the electromagnetic field scattered by the dielectric objects are on the same footing.  This framework was justified on a rigorous theoretical basis by Ciattoni \cite{ciattoni_quantum_2024}, who derived it for finite, dispersive dielectric bodies of arbitrary shape by formulating the theory in the Heisenberg picture within a phenomenological model of dielectric media based on a continuum of harmonic oscillators.   Each elementary region of the dielectric medium is described as a continuum of harmonic oscillators that couples to the electromagnetic field, so that the quantized field experiences the dielectric medium through its macroscopic dielectric permittivity (e.g., \cite{huttner_quantization_1992}, \cite{philbin_canonical_2010}).  As the medium-assisted field, the scattering-assisted field can also be expressed in terms of vacuum fluctuations through the dyadic Green function of dielectric objects \cite{ciattoni_quantum_2024}.

The modified Langevin noise formalism has recently been applied, for the first time, to a single quantum emitter interacting with a linear, dispersive dielectric slab in \cite{na_numerical_2023}, \cite{miano_quantum_2025}, and with a linear, dispersive dielectric sphere in \cite{miano_spectral_2025}. In this framework, the quantum emitter couples to two distinct and independent bosonic reservoirs: the medium-assisted reservoir and the scattering-assisted reservoir. Each reservoir is characterized by its own spectral density, and both spectral densities are quadratic functionals of the dyadic Green function of the dielectric objects, $\mathcal{G}_\omega(\mathbf{r},\mathbf{r}')$. The time evolution of the reduced density operator of the emitter depends on both the initial quantum state of the entire system and on the spectral densities of the two reservoirs. Only under specific conditions can the actions of the two reservoirs be represented by a single equivalent bosonic reservoir whose spectral density depends solely on the value of the dyadic Green function of the dielectric object at the quantum emitter position.
In fact, when the initial quantum state of the entire system is a product state and both reservoirs are initially in the vacuum state, the reduced density operator of the quantum emitter evolves as the emitter interacts with an equivalent single bosonic reservoir, with only positive frequencies and equivalent spectral density $\mathcal{J}(\omega)$ equal to the sum of the spectral density of the individual reservoirs. In this case, one obtains \cite{na_numerical_2023}, \cite{miano_quantum_2025}, \cite{miano_spectral_2025}  $\mathcal{J}(\omega)={\omega^2}{\left(\pi\hbar\varepsilon_0 c^2\right)^{-1}} \mathbf{p}\cdot \operatorname{Im}\left[\mathcal{G}_\omega\left(\mathbf{r}_a, \mathbf{r}_a\right)\right]\cdot\mathbf{p}$ as a consequence of the fundamental integral relation 31 of \cite{ciattoni_quantum_2024}; here $\rb_a$ is the position vector of the emitter and $\mathbf{p}$ is its transition dipole moment. If, instead, both reservoirs are initially in thermal states at the same temperature $T_0$, the reduced dynamics of the quantum emitter can still be recast in terms of a single zero-temperature bosonic effective reservoir with both positive and negative frequencies and a temperature-dependent equivalent spectral density equal to $\theta(\omega,T_0)\mathcal{J}(|\omega|)$ where $\theta(\omega,T_0)=\frac{1}{2}\text{sign}(\omega)\left[1+\coth\left(\frac{\beta_{0}\hbar\omega}{2}\right)\right]$ and $\beta_0=1/(k_BT_0)$. Relations of this type are widely used in the Langevin-
noise literature; however, the conditions under which they remain valid are not always clearly stated. Indeed, when the two reservoirs are prepared at different temperatures, the knowledge of the dyadic Green function of the dielectric objects at the position of the emitter is insufficient to determine the emitter dynamics. Nevertheless, the reduced dynamics of the emitter can still be described by a single bosonic reservoir initially in the vacuum state with an equivalent spectral density equal to the sum of the temperature-dependent medium-assisted  and the temperature-dependent scattering-assisted spectral densities \cite{miano_spectral_2025}. Scenarios in which two reservoirs are prepared in thermal states at different temperatures arise in several contexts \cite{bellomo_creation_2013} and may model different experimental settings, such as cavities driven by thermal photon baths in polaritonic chemistry (e.g., \cite{pannir-sivajothi_blackbody_2025,fassioli_controlling_2025}), and cryogenically cooled cavities (Fabry-Pérot, nanoplasmonic structures, photonic crystal waveguides, etc.) in the presence of thermal photons (e.g. \cite{jarc_cavity-mediated_2023,chiriaco_thermal_2024,bacciconi_dissipation_2025}). In general, if the initial states of the two reservoirs are not thermal, it is no longer possible to represent their action through an equivalent single reservoir.

Although the approach introduced in \cite{miano_quantum_2025}, \cite{miano_spectral_2025} can handle an electromagnetic environment prepared in an arbitrary initial quantum state, it is limited to a single quantum emitter. In the present article, we lift this restriction and generalize this approach based on the modified Langevin noise formalism to a collection of quantum emitters with arbitrary orientations of their transition dipole moments. As we did for the single quantum emitter \cite{miano_quantum_2025}, we apply the emitter-centered mode technique to reduce the number of degrees of freedom of the electromagnetic environment (e.g. \cite{buhmann_casimir-polder_2008}, \cite{feist_macroscopic_2021}).  
In particular, we show that the reduced dynamic of the quantum emitters depends only on the values of the dyadic Green function at the quantum emitter positions whenever the initial state of the entire system is a product state and the initial states of the medium-assisted and the scattering-assisted reservoirs are thermal states with the same temperature. When these conditions are not satisfied, the reduced dynamics of the quantum emitter depends on the generalized spectral densities of the two individual reservoirs, which are quadratic functionals of the dyadic Green function of the dielectric objects.

We proceed as follows. In Section \ref{sec:Model}, we apply the modified Langevin noise formalism to model multiple quantum emitters interacting with a dispersive electromagnetic environment. In Section \ref{sec:EmitterCentered}, we use emitter-centered modes to reduce the number of degrees of freedom of the electromagnetic environment. In Section \ref{sec:Spectral}, we introduce the medium- and the scattering-assisted spectral density matrices. In Section \ref{sec:Correlator}, we propose a surrogate bosonic environment to describe the reduced dynamics of quantum emitters when the initial quantum state of the entire system is a product state and the medium-assisted and scattering-assisted bosonic reservoirs are initially in thermal quantum states with different temperatures. In Section \ref{sec:Application}, we apply our model to a system of two quantum emitters interacting with a nanophotonic environment. We consider two scenarios, a metallic nanosphere and a metallic nanostructure composed of two rods and a disk. Specifically, we explore various aspects of entanglement relaxation, revivals, and generation. In Section \ref{sec:Conclusion}, we give a summary and discuss our conclusions.

\section{Model}
\label{sec:Model}
A collection of $N$ quantum emitters mutually interacts and couples to finite-size dielectric objects of arbitrary shape embedded in an unbounded space. We denote by $\rb_j$ the position vector of the $j$-th quantum emitter with $j=1,2,\ldots, N$. We assume that the dielectrics are linear, isotropic, and dispersive in time. We denote by $V$ the region occupied by the dielectric objects and by  $\varepsilon_\omega(\mathbf{r})$ their relative dielectric permittivity in the frequency domain.  The dispersive dielectrics, together with the electromagnetic field, constitute the electromagnetic environment of the quantum emitters.

\subsection{Hamiltonian}

The Hamiltonian of the entire system, a collection of $N$ quantum emitters plus the electromagnetic environment, is 
\begin{equation}
\label{eq:ham1}
    \hat{H} = \hat{H}_{A} + \hat{H}_{E} + \hat{H}_{I},
\end{equation}
where $\hat{H}_{A}$ is the bare Hamiltonian of the quantum emitters, assumed to be mutually isolated, $\hat{H}_{E}$ is the bare Hamiltonian of the electromagnetic environment and $\hat{H}_{I}$ describes their interaction. 

In the multipolar coupling scheme (Power–Zienau–Woolley picture) and within the dipole approximation, the interaction Hamiltonian $\hat{H}_{I}$ reads
\begin{equation}
  \hat{H}_{I}=-\sum_{j=1}^N\hat{\mathbf{d}}_j\cdot\hat{\mathbf{E}}(\rb_j),
  \label{eq:Hint}
\end{equation}
where $\hat{\mathbf{E}}(\rb)$ is the electric field operator and $\hat{\mathbf{d}}_j$ is the electric dipole moment operator of the $j$-th emitter located at $\rb_j$. We assume that the dipole moment operator of the $j$-th emitter couples through a fixed polarization direction $\mathbf{u}_j$ as in ref. \cite{feist_macroscopic_2021}, so that $\hat{\mathbf{d}}_j=\hat{\mu}_j\mathbf{u}_j$ where $\hat{\mu}_j$ is the corresponding transition dipole moment operator. 
More generally, one may retain up to three orthogonal polarization components per emitter, $\hat{\mathbf{d}}_j=\sum_{a}\hat{\mu}_{ja}\mathbf{a}$ where $a=x,y,z$, and $\hat{\mu}_{jx}, \hat{\mu}_{jy}, \hat{\mu}_{jz}$ are the corresponding transition-dipole moment operators.

\subsection{ Diagonal Form of $\hat{H}_E$}

The bare Hamiltonian of the electromagnetic environment $\hat{H}_{E}$  accounts for the electromagnetic field, the polarization currents of the dielectric objects, and their interaction. The modified Langevin noise formalism provides a straightforward route to diagonalize it \cite{ciattoni_quantum_2024}. 

We express the electric field operator in the Schr{\"o}dinger picture as $\hat{\mathbf{E}}(\rb)=\int^\infty_{0} d\omega [\hat{\mathbf{E}}_\omega(\rb) + h.c.] $ where $\hat{\mathbf{E}}_\omega(\rb)$ is its monochromatic component in the Heisenberg picture. Within the modified Langevin noise formalism, the electric field operator $\hat{\mathbf{E}}_\omega(\rb)$ has two contributions, the medium-assisted contribution $\hat{\mathbf{E}}^{(M)}_\omega(\rb)$ and the scattering-assisted contribution $\hat{\mathbf{E}}^{(S)}_\omega(\rb)$,
\begin{equation}
\label{eq:Elect}
    \hat{\mathbf{E}}_\omega = \hat{\mathbf{E}}_\omega^{(M)} + \hat{\mathbf{E}}_\omega^{(S)}.
\end{equation}
The medium-assisted contribution is generated by the noise polarization currents of the dielectric object \cite{gruner_green-function_1996}, while the scattering-assisted contribution is generated by the vacuum fluctuations of the electromagnetic field scattered by the dielectric object \cite{stefano_mode_2001}, \cite{drezet_quantizing_2017}. These two contributions are expressed in terms of particular bosonic operators that diagonalize $\hat{H}_{E}$.

The medium-assisted field $\hat{\mathbf{E}}_\omega^{(M)}(\rb)$ is given by \cite{ciattoni_quantum_2024}
\begin{equation}
\hat{\mathbf{E}}_\omega^{(M)}(\mathbf{r})= \int_V d^3 \mathbf{r}^{\prime} \, \mathcal{G}_{e}\left(\mathbf{r}, \mathbf{r}^{\prime};\omega\right) \hat{\mathbf{f}}_{\omega}\left(\mathbf{r}^{\prime}\right),
\label{eq:Emat}
\end{equation}
where $\hat{\mathbf{f}}_{\omega}\left(\mathbf{r}\right)$ is the monochromatic bosonic field operator describing the fluctuations of the dielectric polarization currents, whose support is the region $V$,
\begin{equation}
\mathcal{G}_{e}\left(\mathbf{r}, \mathbf{r}^{\prime}; \omega\right)= i \frac{\omega^2}{c^2} \sqrt{\frac{\hbar}{\pi \varepsilon_0} \operatorname{Im}\left[\varepsilon_\omega\left(\mathbf{r}^{\prime}\right)\right]} \mathcal{G}_\omega\left(\mathbf{r}, \mathbf{r}^{\prime}\right),
\label{eq:Ge}
\end{equation}
$\mathcal{G}_\omega\left(\mathbf{r}, \mathbf{r}^{\prime}\right)$ is the dyadic Green's function of the dielectric object satisfying the
equation
\begin{equation}
\left(\nabla_\rb \times \nabla_\rb \times-k_\omega^2 \varepsilon_\omega\right) \mathcal{G}_\omega\left(\mathbf{r}, \mathbf{r}^{\prime}\right)=\delta\left(\mathbf{r}-\mathbf{r}^{\prime}\right) \mathcal{I},
\end{equation}
and the boundary condition $\mathcal{G}_\omega\left(\mathbf{r}, \mathbf{r}^{\prime}\right) \rightarrow 0$ for $r, r^{\prime} \rightarrow$ $\infty$; $\varepsilon_0$ is the vacuum permittivity, $k_\omega=\omega/c$, $c$ is the light velocity in vacuum, and $ \mathcal{I}$ is the identity dyad. 

Let $\mathbf{F}_{\omega \mathbf{n} \nu}(\mathbf{r})$ denote the solution of a homogeneous equation 
\begin{equation}
\left( \nabla_\rb \times \nabla_\rb \times-k_\omega^2 \varepsilon_\omega\right) \mathbf{F}_{\omega \mathbf{n} \nu}=0,
\label{eq:Ffield}
\end{equation}
satisfying the asymptotic boundary condition
\begin{equation}
\mathbf{F}_{\omega \mathbf{n} \nu}(\mathbf{r}) \underset{r \rightarrow \infty}{\approx} e^{ik_{\omega}\rb\cdot \mathbf{n}}\mathbf{e}_{\mathbf{n} \nu},
\label{eq:FfieldBC}
\end{equation}
where $\mathbf{n}$ is the unit vector along the wave vector $\mathbf{k} = k_{\omega}\mathbf{n}$ and $\mathbf{e}_{\mathbf{n}1}$, $\mathbf{e}_{\mathbf{n}2}$ are two mutually orthogonal polarization unit vectors orthogonal to $\mathbf{n}$. We introduce the scattering modes $\mathbf{E}_{\omega \mathbf{n} \nu}(\mathbf{r})$ defined as \cite{ciattoni_quantum_2024}
\begin{equation}
\mathbf{E}_{\omega \mathbf{n} \nu}(\mathbf{r}) =\sqrt{\frac{\hbar \omega^3}{16 \pi^3 \varepsilon_0 c^3}} \mathbf{F}_{\omega \mathbf{n} \nu}(\mathbf{r}).
\label{eq:ScatteringModes}
\end{equation}
The scattering-assisted field $\hat{\mathbf{E}}_\omega^{(S)}$ is given by
\begin{equation}
\hat{\mathbf{E}}^{(S)}_\omega(\mathbf{r})=
\oint d o_{\mathbf{n}} \sum_\nu \mathbf{E}_{\omega \mathbf{n} \nu}(\mathbf{r}) \hat{g}_{\omega \mathbf{n} \nu},
\label{eq:Esca}
\end{equation}
where $\hat{g}_{\omega \mathbf{n} \nu}$ is the monochromatic bosonic operator that describes the fluctuation of the radiation incoming from infinity and scattered by the dielectric object. Here $o_{\mathbf{n}}=(\theta_{\mathbf{n}},\phi_{\mathbf{n}})$ are the polar angles of the unit vector ${\mathbf{n}}$, $do_{\mathbf{n}}=\sin\theta_{\mathbf{n}} d\theta_{\mathbf{n}} d\phi_{\mathbf{n}}$ is the differential of the solid angle and the integration is over the full solid angle, $\theta \in [0, \pi]$ and $\phi \in [0, 2 \pi]$.

The bosonic field operators $\hat{\mathbf{f}}_{\omega}(\mathbf{r})$ and $\hat{g}_{\omega \mathbf{n} \nu}$ are independent. Any possible commutation relations between them vanishes except the fundamental ones,
\begin{subequations}
\begin{align}
   {\left[\hat{\mathbf{f}}_{\omega}(\mathbf{r}), \hat{\mathbf{f}}_{\omega^{\prime}}^{\dagger}\left(\mathbf{r}^{\prime}\right)\right] } &=\delta\left(\omega-\omega^{\prime}\right) \delta\left(\mathbf{r}-\mathbf{r}^{\prime}\right) \mathcal{I}, \\
   {\left[\hat{g}_{\omega \mathbf{n} \nu}, \hat{g}_{\omega^{\prime} \mathbf{n}^{\prime} \nu^{\prime}}^{\dagger}\right] } &=\delta\left(\omega-\omega^{\prime}\right) \delta\left(o_{\mathbf{n}}-o_{\mathbf{n}^{\prime}}\right) \delta_{\nu \nu^{\prime}}, 
\end{align}    
\end{subequations}
where $\delta\left(o_{\mathbf{n}}-o_{\mathbf{n}^{\prime}}\right)=\delta\left(\theta_{\mathbf{n}}-\theta_{\mathbf{n}}^{\prime}\right) \delta\left(\phi_{\mathbf{n}}-\phi_{\mathbf{n}}^{\prime}\right) / \sin \theta_{\mathbf{n}}$.
Together with the expressions of the medium-assisted electric field $\hat{\mathbf{E}}^{(M)}_\omega$ and of the scattering-assisted electric field $\hat{\mathbf{E}}^{(S)}_\omega$,  these commutation relations ensure the canonical commutation relations for the electromagnetic field and the bath oscillator fields that describe dielectric objects \cite{ciattoni_quantum_2024}. 

The bosonic field operators $\hat{\mathbf{f}}_{\omega}(\mathbf{r})$ and $\hat{g}_{\omega \mathbf{n} \nu}$ diagonalize the electromagnetic environment Hamiltonian,
\begin{equation}
\label{eq:HEtot}
\hat{H}_{E}= \hat{H}^{(M)}_E+ \hat{H}^{(S)}_E,
\end{equation}
where
\begin{subequations}
\begin{align}
  \hat{H}^{(M)}_E=\int_0^{\infty} d \omega \hbar \omega \int_V d^3 \, \mathbf{r} \,\hat{\mathbf{f}}_{\omega}^{\dagger}(\rb) \cdot \hat{\mathbf{f}}_{\omega}(\rb), \\
   \hat{H}^{(S)}_E=\int_0^\infty d\omega \hbar\omega\, \oint d o_{\mathbf{n}} \sum_\nu \hat{g}_{\omega \mathbf{n} \nu}^{\dagger} \hat{g}_{\omega \mathbf{n} \nu}, 
\end{align}    
\end{subequations}
are, respectively, the contribution of the medium and scattering-assisted fields.
The operators $\{\hat{\mathbf{f}}_{\omega}^{\dagger}$, $\hat{\mathbf{f}}_{\omega}\}$ and $\{\hat{g}_{\omega \mathbf{n} \nu}^{\dagger}$, $\hat{g}_{\omega \mathbf{n} \nu}\}$ are the creation and annihilation operators of polaritonic excitations in two independent bosonic reservoirs, the medium-assisted and scattering-assisted bosonic reservoirs, respectively.

The fundamental integral identity \cite{na_numerical_2023,ciattoni_quantum_2024}
\begin{equation}
\label{eq:sum}
\mathcal{A}_\omega(\rb,\rb')+ \mathcal{B}_\omega(\rb,\rb')=\frac{\hbar \omega^2}{\pi \varepsilon_0 c^2} \operatorname{Im}\left[\mathcal{G}_\omega\left(\mathbf{r}, \mathbf{r}^{\prime}\right)\right]
\end{equation}
holds, where 
\begin{subequations}
\begin{align}
   \mathcal{A}_\omega(\rb,\rb')&=\int_V d^3 \mathbf{r''} \, \mathcal{G}_{e}(\mathbf{r}, \mathbf{r}''; \omega) \cdot \mathcal{G}_{e}^{* T}\left(\mathbf{r}^{\prime}, \mathbf{r''};\omega\right), \\
    \mathcal{B}_\omega(\rb,\rb')&=\oint d o_{\mathbf{n}} \sum_\nu \mathbf{E}_{\omega \mathbf{n} \nu}(\mathbf{r}) \mathbf{E}_{\omega \mathbf{n} \nu}^*\left(\mathbf{r}^{\prime}\right).
\end{align}    
\end{subequations}
 This identity was first validated numerically by D.~Y.~Na and coworkers \cite{na_numerical_2023}. 

From this property it follows that
\begin{equation}
{\left[\hat{\mathbf{E}}_\omega(\mathbf{r}), \hat{\mathbf{E}}_{\omega^{\prime}}^{\dagger}\left(\mathbf{r}^{\prime}\right)\right]}=\frac{\hbar \omega^2}{\pi\varepsilon_0 c^2} \operatorname{Im}\left[\mathcal{G}_\omega\left(\mathbf{r}, \mathbf{r}^{\prime}\right)\right]\delta\left(\omega-\omega^{\prime}\right),
\end{equation}
which is the standard electric field commutator expressed in terms of the dyadic Green function of the dielectric.

\section{Bright modes of the electromagnetic environment and reduced Hamiltonian}
\label{sec:EmitterCentered}
We now represent the vector field operators $\{\hat{\mathbf{f}}_{\omega}(\mathbf{r})\}$ and the scalar field operators $\{\hat{g}_{\omega \mathbf{n} \nu } \}$ in terms of the emitter-centered modes (e.g. \cite{buhmann_casimir-polder_2008}, \cite{feist_macroscopic_2021}, \cite{miano_quantum_2025}).
Following Ref.~\cite{feist_macroscopic_2021}, we aim to eliminate the degrees of freedom of the medium-assisted reservoir and the scattering-assisted reservoir that are not excited by the quantum emitters. Hence, starting from Eqs. \ref{eq:Hint}, \ref{eq:Elect}, and \ref{eq:Emat}, we introduce the auxiliary interaction operator $\hat{F}_j^{(M)}$ describing the coupling between the $j$-th emitter and the medium-assisted reservoir
\begin{equation}
\hat{F}_j^{(M)}= \int_0^\infty d\omega \int_{V}d^3 \mathbf{r}\, [\mathbf{u}_j \cdot \mathcal{G}_{e}(\mathbf{r}_j, \mathbf{r}, \omega)\cdot \hat{\mathbf{f}}_{\omega}(\mathbf{r})+h.c.].
\label{eq:FM}
\end{equation}
Similarly, from Eqs. \ref{eq:Hint}, \ref{eq:Elect}, and \ref{eq:Esca}, we introduce the auxiliary interaction operator $\hat{F}_{j}^{(S)}$, describing the coupling between the $j$-th emitter and the scattering-assisted reservoir
\begin{equation}
\hat{F}_{j}^{(S)}= \int_0^\infty d\omega \oint d{o}_{\mathbf{n}}\, [\sum _\nu \, \mathbf{u}_j \cdot \mathbf{E}_{\omega \mathbf{n} \nu}(\rb_j) \hat{g}_{\omega \mathbf{n} \nu }+h.c.] \, .
\label{eq:FS}
\end{equation}
The interaction Hamiltonian \ref{eq:Hint} can then be recast as 
\begin{equation}
\hat{H}_I=-\sum_{j=1}^N\hat{\mu}_j\hat{F}_j,
\end{equation}
where
\begin{equation}
\hat{F}_j=\hat{F}_j^{(M)}+\hat{F}_j^{(S)}.
\end{equation}

\subsection{Medium-Assisted Reservoir}
\label{sec:BrightDark}

The structure of the expression \ref{eq:FM} of $\hat{F}^{(M)}_j$  suggests introducing the monochromatic scalar bosonic operator
\begin{equation}
\hat{A}_j(\omega)=\int_{V}d^3 \mathbf{r}\, \boldsymbol{\alpha}_j(\rb;{\omega})\cdot \hat{\mathbf{f}}_{\omega}{ (\mathbf{r})},
\end{equation}
with $j=1,\ldots,N$, where
\begin{equation}
\label{eq:alpha}
\boldsymbol{\alpha}_{j}(\rb;{\omega})=\mathbf{u}_j \cdot \mathcal{G}_{e}(\mathbf{r}_j, \mathbf{r};\omega){,}
\end{equation}
so that
\begin{equation}
\hat{F}^{(M)}_j=\int_0^\infty d\omega [\hat{A}_j(\omega) + h.c.] \, .
\end{equation}
The operators $\{\hat{A}_{i}(\omega)\}$ obey
\begin{equation}
{\left[\hat{A}_{i}(\omega), \hat{A}_{j}^{\dagger}(\omega^{\prime})\right] } =M_{ij}(\omega)\delta\left(\omega-\omega^{\prime}\right),
\end{equation}
where
\begin{equation}
   M_{ij}(\omega)= \int_V d^3\mathbf{r}\,  \boldsymbol{\alpha}_i(\rb;{\omega}) \cdot \boldsymbol{\alpha}_j^*(\rb;{\omega}).
\end{equation}
The $N\times N$ matrix $M=[M_{ij}]$ is a complex Gram matrix associated  with the inner product $\int_Vd\rb^3 \mathbf{a}(\rb) \cdot \mathbf{b}^*(\rb)$. By construction $M$ is Hermitian. The vector fields $\{\boldsymbol{\alpha}_{j}(\rb;\omega)\}$ are linearly independent, and then $M$ is positive definite. Since these fields are not mutually orthogonal with respect to the inner product above, $M$ is, in general, dense. Using \ref{eq:alpha}, we obtain
\begin{equation}
\label{eq:M}
M_{ij}(\omega)=\int_V d^3 \mathbf{r} \, \mathbf{u}_i\cdot \mathcal{G}_{e}(\rb_i,\rb ;\omega) \mathcal{G}_{e}^{* T}(\rb_j,\rb;\omega) \cdot \mathbf{u}_j.
\end{equation}

We now seek a representation of $\hat{H}_{E}^{(M)}$ in terms of the scalar operators $\{\hat{A}_j(\omega)\}$.  The vector fields $\{\boldsymbol{\alpha}_j(\rb;\omega)\}$ are, in general, not orthogonal with respect to the above inner product; consequently, the basis $\{\hat{A}_j(\omega)\}$  does not diagonalize $\hat{H}_{E}^{(M)}$. To obtain a diagonal form, we generate the orthonormal set of vector fields $\{\boldsymbol{\chi}_i(\rb;\omega)\}$ by applying the linear transformation
\begin{equation}
\boldsymbol{\chi}_i(\rb;\omega)= \sum_{j=1}^N U_{ij}(\omega)\boldsymbol{\alpha}_j(\rb;\omega),
\end{equation}
where the $N \times N$ matrix $U=[U_{ij}]$ is such that $ UMU^{\dagger}=I $ and $I$ is the identity matrix. Completing $\{\boldsymbol{\chi}_i(\rb;\omega)\}$ with an orthonormal set $\{\boldsymbol{\chi}^{\mathrm{dark}}_m(\rb;\omega)\}$, we express the field operator $\hat{\mathbf{f}}_{\omega}(\mathbf{r})$ as
\begin{equation}
\label{eq:espf}
\hat{\mathbf{f}}_{\omega}(\mathbf{r})=\sum_{j=1}^N \boldsymbol{\chi}_j^*(\rb;\omega) \hat{C}_j(\omega) + \sum_m [\boldsymbol{\chi}_m^\text{dark}(\rb;\omega)]^*\hat{C}_m^\text{dark}(\omega),
\end{equation}
where the new scalar bosonic operator $\hat{C}_i(\omega)$ is given by
\begin{equation}
\hat{C}_i(\omega)= \sum_{j=1}^N U_{ij}(\omega)\hat{A}_j(\omega).
\end{equation}
Note that every vector field $\boldsymbol{\chi}_m^\text{dark}(\rb;\omega)$ does not couple to the emitters: the vector fields $\{\boldsymbol{\chi}_i(\rb;\omega)\}$ are the bright modes of the medium assisted field, while the vector fields $\{\boldsymbol{\chi}_m^\text{dark}(\rb;\omega)\}$ are the dark modes. The scalar bosonic operators $\hat{C}_i(\omega)$ are the medium-assisted bright bosonic operators of the system and $\hat{C}_m^\text{dark}(\omega)$ are the dark bosonic operators. The matrix $U$ can be chosen in various ways \cite{feist_macroscopic_2021}. For our purposes, it is convenient to choose  $ U=M^{-1/2}$.

Monochromatic bosonic operators $\{\hat{C}_i(\omega)\}$ satisfy
\begin{equation}
\label{eq:commut_C}
{\left[\hat{C}_{i}(\omega), \hat{C}_{j}^{\dagger}(\omega^{\prime})\right] } =\delta_{ij} \delta\left(\omega-\omega^{\prime}\right).
\end{equation}
The bosonic operators $\hat{A}_i(\omega)$ are related to $\hat{C}_i(\omega)$ through the equations
\begin{equation}
\hat{A}_i(\omega)= \sum_{j=1}^N G^{(M)}_{ij}(\omega)\hat{C}_j(\omega),
\end{equation}
where the $N \times N$ matrix $G^{(M)}=[G^{(M)}_{ij}]$ is the inverse of the matrix $U$,
$ G^{(M)}= {M}^{1/2}$.
Using this relation, we express $\hat{F}^{(M)}_i$ through the bright bosonic operators $\{\hat{C}_j\}$,
\begin{equation}
\label{eq:F_M}
\hat{F}_i^{(M)}=\sum_{j=1}^N \int_0^\infty d\omega [G_{ij}^{(M)}(\omega)\hat{C}_j(\omega)+h.c.].
\end{equation}
Using \ref{eq:espf}, we obtain for $\hat{H}_{E}^{(M)}$
\begin{multline}
\hat{H}_{E}^{(M)}=\sum_{j=1}^N \int_0^\infty d\omega \hbar \omega \hat{C}_j ^\dagger(\omega) \hat{C}_j(\omega) \\ + \sum_m \int_0^\infty d\omega \hbar \omega  \hat{C}_m^{dark\dagger}({\omega}) \hat{C}_m^\text{dark}({\omega}).
\end{multline}

\subsection{Scattering-Assisted Reservoir}

As in the medium-assisted case, the expression \ref{eq:FS} of $\hat{F}_{j}^{(S)}$ suggests introducing the scalar monochromatic bosonic operator
\begin{equation}
\hat{B}_j (\omega)= \oint d o_{\mathbf{n}} \sum _\nu \, {\beta}_j (\mathbf{n}, \nu;\omega) \hat{g}_{\omega \mathbf{n} \nu },
\end{equation}
with $j=1,\ldots,N$, where
\begin{equation}
{\beta}_j (\mathbf{n}, \nu;\omega) =\mathbf{u}_j \cdot \mathbf{E}_{\omega \mathbf{n} \nu}(\mathbf{r}_j),
\end{equation}
so that
\begin{equation}
\hat{F}^{(S)}_j=\int_0^\infty d\omega [\hat{B}_j(\omega) + h.c.] \, .
\end{equation}
The operators $\{\hat{B}_i  (\omega)\}$  obey
\begin{equation}
{\left[\hat{B}_{i}(\omega), \hat{B}_{j}^{\dagger}(\omega^{\prime})\right] } =S_{ij}(\omega)\delta\left(\omega-\omega^{\prime}\right),
\end{equation}
where
\begin{equation}
\label{eq:S}
S_{ij}(\omega)= \oint do_{\mathbf{n}} \, \mathbf{u}_i \cdot \sum_\nu \mathbf{E}_{\omega \mathbf{n} \nu}(\mathbf{r}_i) \mathbf{E}_{\omega \mathbf{n} \nu}^*(\mathbf{r}_j) \cdot \mathbf{u}_j.
\end{equation}
The $N \times N$ matrix $S=[S_{ij}]$ is also a complex Gram matrix with respect to the scalar product $\int do_{\mathbf{n}} \sum_\nu {f}( \mathbf{n},\nu) {g}^*( \mathbf{n},\nu) $. Therefore, $S$ is Hermitian by construction; since the vectors $\{\beta_j(\mathbf{n},\nu,\omega)\}$ are linearly independent, it is also positive definite. Proceeding as in the medium-assisted case, we introduce the set of  functions $ \left\{ \xi_j(\mathbf{n},\nu;\omega) \right\}$, orthonormal with respect to the above scalar product.  They are given by
\begin{equation}
\label{eq:scat}
\xi_i(\mathbf{n},\nu;\omega)= \sum_{j=1}^N V_{ij}{\beta}_j (\mathbf{n}, \nu;\omega),
\end{equation}
where the $N \times N$ matrix $V=[V_{ij}]$ is such that
$VSV^{\dagger}=I$.
In analogy with the medium-assisted case, we choose  
$ V=S^{-1/2}$. Completing the set $\{{\xi}_i(\mathbf{n},\nu;\omega)\}$ with an orthonormal set $\{{\xi}^{\mathrm{dark}}_m\}$, we express the field operator $\hat{g}_{\omega \mathbf{n} \nu}$ as
\begin{equation}
\hat{g}_{\omega \mathbf{n} \nu}= \sum_{j=1}^N\xi_j^*(\mathbf{n},\nu, \omega) \hat{D}_j(\omega) + \sum_m [{\xi}_m^\text{dark}]^* \hat{D}_m^\text{dark}(\omega),
\end{equation}
where the bosonic operators $\{\hat{D}_i(\omega)\}$ are given by
\begin{equation}
\hat{D}_i(\omega)= \sum_{j=1}^N V_{ij}(\omega)\hat{B}_j(\omega).
\end{equation}
As for the medium assisted case, the scalar fields $\{\xi_m^\text{dark}\}$ do not couple to the emitters; 
$\{\xi_i\}$ are the bright modes of the scattering-assisted field and $\{\hat{D}_i\}$ are the corresponding bright operators.
The scalar bosonic operators $\{\hat{B}_i(\omega)\}$ are related to $\{\hat{D}_i(\omega)\}$ through the equations
\begin{equation}
\hat{B}_i(\omega)= \sum_{j=1}^N G_{ij}^{(S)} (\omega)\hat{D}_j(\omega),
\end{equation}
where the $N \times N$ matrix $G^{(S)}=[G^{(S)}_{ij}]$ is the inverse of the matrix $V$,
$ G^{(S)}=S^{1/2}$. Using this relation, we express $\hat{F}_{i}^{(S)}$ through the bright bosonic operators $\{\hat{D}_j(\omega)\}$,
\begin{equation}
\hat{F}_{i}^{(S)}=\sum_{j=1}^N \int_0^\infty d\omega [G_{ij}^{(S)}(\omega)\hat{D}_j(\omega)+h.c.].
\end{equation}
Using \ref{eq:scat}, we obtain for $\hat{H}_{E}^{(S)}$
\begin{multline}
\hat{H}_{E}^{(S)}=\sum_{j=1}^N \int_0^\infty d\omega \hbar \omega \hat{D}_j ^\dagger(\omega) \hat{D}_j(\omega) \\ + \sum_m \int_0^\infty d\omega \hbar \omega \hat{D}_m^{dark\dagger}({\omega}) \hat{D}_m^\text{dark}({\omega}).
\end{multline}

\subsection{Reduced Hamiltonian}
\label{sec:MinimalRep}

The medium-assisted and scattering-assisted dark modes are decoupled from the rest of the system, do not affect the dynamics of the quantum emitters, and can be dropped from the expressions of $\hat{H}_{E}^{(M)}$ and $\hat{H}_{E}^{(S)}$. 
The overall system, i.e., quantum emitters + bright modes of the electromagnetic environment, is described by the reduced Hamiltonian
\begin{equation}
\label{eq:Hbright}
\hat{H}_{red} = \hat{H}_{A} + \hat{H}_{E}^\text{bright} + \hat{H}_{I},
\end{equation}
where
\begin{equation}
   \hat{H}_{E}^\text{bright}=\sum_{i=1}^N \int_0^\infty d\omega \hbar \omega \left[\hat{C}_i ^\dagger(\omega) \hat{C}_i(\omega) +\hat{D}_i ^\dagger(\omega) \hat{D}_i(\omega)\right] 
   \label{eq:HEbright},
\end{equation}
and
\begin{multline}\label{eq:Hintbright}
  \hat{H}_{I}=-\sum_{i,j=1}^N \hat{\mu}_i \left[ \int_0^\infty d\omega G_{ij}^{(M)}(\omega)\hat{C}_j(\omega)+h.c.\right] \\
-\sum_{i,j=1}^N \hat{\mu}_i \left[ \int_0^\infty d\omega G_{ij}^{(S)}(\omega)\hat{D}_j(\omega)+h.c.\right].  
\end{multline}
In summary, the set of quantum emitters behaves as an open quantum system coupled to two independent reservoirs: the medium-assisted reservoir and the scattering-assisted reservoir. Each reservoir consists of $N$ independent set of bosonic modes. The medium-assisted reservoir is characterized by the coupling matrix ${G}^{(M)}(\omega)=M(\omega)^{1/2}$, and the scattering-assisted reservoir by ${G}^{(S)}(\omega)=S(\omega)^{1/2}$. In Appendix \ref{sec:Bright}, we provide the expression of the electric field operator due to the bright modes.

\section{Spectral density matrices}
\label{sec:Spectral}
The matrices $G^{(M)}(\omega)$ and $G^{(S)}(\omega)$ fully characterize the interaction of the quantum emitters with the electromagnetic environment, including their mutual interactions. It is convenient to recast these coupling matrices in terms of two matrices that generalize the concepts of medium-assisted spectral density and scattering-assisted spectral density, as originally introduced in \cite{miano_quantum_2025} and \cite{miano_spectral_2025} for a single quantum emitter, to configurations of multiple quantum emitters.

Let us introduce the $N\times N$ diagonal matrix $P$ of the transition dipole moments of the quantum emitters as
\begin{equation}
    P= \text{diag}(\mu_1,\mu_2,\ldots,\mu_N),
\end{equation}
where $\mu_j$ is the transition dipole moment of the $j$-th quantum emitter. The \textit{medium-assisted spectral density matrix} $\mathcal{J}^{(M)}$ and the \textit{scattering-assisted spectral density matrix} $\mathcal{J}^{(S)}$ are defined as
\begin{subequations}
\begin{align}
    \mathcal{J}^{(M)}(\omega)=\frac{1}{\hbar^2} P \, [{G}^{(M)}(\omega)]^2 \, P,\\
   \mathcal{J}^{(S)}(\omega)=\frac{1}{\hbar^2} P \, [{G}^{(S)}(\omega)]^2 \, P,
\end{align}
\end{subequations}
so that, equivalently,
\begin{subequations}
\begin{align}
   {G}^{(M)}(\omega)&= \hbar [P^{-1}\mathcal{J}^{(M)}(\omega)P^{-1}]^{1/2},\\
   {G}^{(S)}(\omega)&= \hbar [P^{-1}\mathcal{J}^{(S)}(\omega)P^{-1}]^{1/2}.
\end{align}
\end{subequations}
Since $G^{(M)}=M^{1/2}$ and $G^{(S)}=S^{1/2}$, we obtain
\begin{subequations}
\begin{align}
    \mathcal{J}_{i j}^{(M)}(\omega)&= \frac{\mu_i \mu_j}{\hbar^2}M_{ij},\\
    \mathcal{J}_{i j}^{(S)}(\omega)&= \frac{\mu_i \mu_j}{\hbar^2}S_{ij},
\end{align}
\end{subequations}
and using \ref{eq:M} and \ref{eq:S},
\begin{subequations}
\begin{align}
    \label{eq:Jmat}
    \mathcal{J}_{i j}^{(M)}(\omega)&= \frac{\mu_i \mu_j}{\hbar^2} \int_V d^3 \mathbf{r} \, \mathbf{u}_i\cdot \mathcal{G}_{e}(\rb_i,\rb) \mathcal{G}_{e}^{* T}(\rb_j,\rb) \cdot \mathbf{u}_j, \\
    \label{eq:Jsca}
    \mathcal{J}_{i j}^{(S)}(\omega)&=\frac{\mu_i \mu_j}{\hbar^2}\oint do_{\mathbf{n}} \, \sum_\nu \mathbf{u}_i \cdot \mathbf{E}_{\omega \mathbf{n} \nu}(\mathbf{r}_i) \mathbf{E}_{\omega \mathbf{n} \nu}^*(\mathbf{r}_j) \cdot \mathbf{u}_j.
\end{align}
\end{subequations}
Both spectral density matrices are complex, Hermitian, and positive definite. The elements $\mathcal{J}_{i j}^{(M)}(\omega)$ and $\mathcal{J}_{i j}^{(S)}(\omega)$ are not independent; in fact, as a consequence of \ref{eq:sum}
\begin{equation}
\label{eq:Fr}
\mathcal{J}_{i j}^{(M)}(\omega)+\mathcal{J}_{i j}^{(S)}(\omega)=\frac{1}{2\pi} \Gamma_{ij}(\omega),
\end{equation}
where
\begin{equation}
\label{eq:gequiv1}
\Gamma_{ij}(\omega) = \frac{2\omega^2}{\hbar\varepsilon_0 c^2} (\mu_i\mathbf{u}_i)\cdot \operatorname{Im}\left[\mathcal{G}_\omega\left(\mathbf{r}_i,\mathbf{r}_j\right)\right]\cdot (\mu_j\mathbf{u}_j).
\end{equation}
The matrix $\Gamma=[\Gamma_{ij}]$ is real, symmetric, and positive definite.

The medium-assisted spectral density and the scattering-assisted spectral density have a direct interpretation within classical electrodynamics. Consider the classical electromagnetic field generated, in the presence of dielectric objects, by $N$ electric dipoles with dipole moments $\mu_j \mathbf{u}_j$, located at ${\mathbf{r}}_j$ with $j=1,\ldots,N$ and oscillating at frequency $\omega$. 
We find that $\frac{\pi}{2}\hbar \omega \sum_{i,j=1}^N \mathcal{J}^{(M)}_{ij}(\omega)$ is equal to the average electromagnetic power absorbed by the dielectric bodies; $\frac{1}{4} (\hbar \omega)\sum_{i,j=1}^N \Gamma_{ij}(\omega)$ is equal to the averaged electromagnetic power emitted by the set of electric dipoles. Then $\frac{\pi}{2}\hbar \omega \sum_{i,j=1}^N \mathcal{J}^{(S)}_{ij}(\omega)$ is equal to the average electromagnetic power radiated toward infinity. Accordingly, Eq. \ref{eq:Fr} represents the statement of the Poynting theorem for the investigated scenario. See Appendix~C for a full discussion.

\section{Correlator matrices and surrogate environment}
\label{sec:Correlator}
The dynamics of $N$ quantum emitters coupled to $N$ medium-assisted bosonic modes characterized by the spectral density matrix $\mathcal{J}^{(M)}(\omega)$ and to $N$ scattering-assisted bosonic modes characterized by the spectral density matrix $\mathcal{J}^{(S)}(\omega)$ can, in principle, be computed numerically with a variety of methods \cite{schollwock_density-matrix_2011,medina_few-mode_2021}. However, this is often computationally demanding. Under specific conditions, analogous to the single quantum emitter case  \cite{miano_quantum_2025}, \cite{miano_spectral_2025}, the actions of the $2N$ bosonic modes can be represented by only $N$ bosonic modes characterized by an equivalent $N\times N$ spectral density matrix, in the same spirit as in \cite{tamascelli_nonperturbative_2018,tamascelli_efficient_2019}.

We assume that: (i) the quantum state of the composite system at the initial time $t_0=0$ is a product state, that is, $\hat{\rho}(0)=\hat{\rho}_A(0) \otimes \hat{\rho}_E^{(M)}(0) \otimes \hat{\rho}_E^{(S)}(0)$, where $\hat{\rho}(0)$, $\hat{\rho}_A (0)$, $\hat{\rho}_E^{(M)}(0)$ and $\hat{\rho}_E^{(S)}(0)$ are the initial density operators of the entire system, the set of quantum emitters, the medium-assisted reservoir and the scattering-assisted reservoir, respectively; (ii) the initial states of both reservoirs are Gaussian. Then, the evolution of the reduced density operator of the set of quantum emitters $\hat{\rho}_A(t) = \text{Tr}_E \left[ \hat{\rho} (t) \right]$ depends only on the expectation values ${F}_j(t)$ 
and the two-time correlation functions ${C}_{ij}(t+\tau;t)$ (e.g. \cite{h_p_breuer_and_f_petruccione_theory_2002}) of the environment interaction operators $\hat{F}_j$ with $i,j=1, 2, \ldots, N$,
\begin{subequations}
\begin{align}
  {F}_j(t) &= \text{Tr}_E \left[\hat{F}_j(t) \hat{\rho}_{E}(0) \right], \\
   {C}_{ij}(t+\tau;t)&= \text{Tr}_E \left[{\hat{F}}_i(t+\tau){\hat{F}}_j(t) \hat{\rho}_{E}(0) \right],
\end{align}    
\end{subequations}
where $\hat{\rho}_E(0)\equiv \hat{\rho}_E^{(M)}(0) \otimes \hat{\rho}_E^{(S)}(0)$,
\begin{equation}
\hat{{F}}_j(t) = \hat{U}^\dagger_E(t)\hat{F}_j \hat{U}_E(t),
\end{equation}
and $\hat{U}_E(t)=\exp(-i\hat{H}_{E}^\text{bright}t/\hbar)$ is the free evolution operator of the bright bosonic modes of the electromagnetic environment. Using this property, it is possible to introduce an equivalent \textit{surrogate} environment with only $N$ bosonic modes that reproduces the same reduced dynamics of the quantum emitters.

The expectation values and the two-time correlation functions of the interaction operators are given by:
\begin{equation}
 {F}_j(t) = {F}^{(M)}_j(t) + {F}^{(S)}_j(t),
\end{equation}
and
\begin{multline}
   C_{ij}(t+\tau;t) = C^{(M)}_{ij}(t+\tau;t) + C^{(S)}_{ij}(t+\tau;t)+ \\ 
   {F}^{(M)}_i(t+\tau) {F}^{(S)}_j(t)+ {F}^{(S)}_i(t+\tau) {F}^{(M)}_j(t),   
\end{multline}
where ${F}^{(\alpha)}_i(t)=\text{Tr}_E \left[\hat{F}^{(\alpha)}_i(t) \hat{\rho}_{E}(0) \right]$ and ${C}^{(\alpha)}_{ij}(t+\tau;t)= \text{Tr}_E \left[{\hat{F}}^{(\alpha)}_i(t+\tau){\hat{F}}^{(\alpha)}_j(t) \hat{\rho}_{E}(0) \right]$ with $\alpha=M, S $. If these expectation values vanish, we have
\begin{equation}
C_{ij}  = C^{(M)}_{ij}  + C^{(S)}_{ij} .
\end{equation}
This is the case, for instance, when both reservoirs are initially in vacuum, thermal, or undisplaced squeezed states. In such situations, the  influence of the electromagnetic environment on the emitter ensemble is fully characterized by the $N\times N$ correlator matrix $C(t+\tau;t)=[C_{ij}(t+\tau;t)]$.

\subsection{Reservoirs Initially in Thermal Quantum States}

We now consider scenarios in which the two reservoirs are initially in thermal quantum states at temperatures $T_0^{(M)}$ and $T_0^{(S)}$.  The corresponding correlators are then stationary, i.e.,
$C_{ij}^{(\alpha)}(\tau)\equiv C_{ij}^{(\alpha)}(t+\tau;t)$ for $\alpha=M,S$, and read (see Appendix~\ref{sec:Correlators}):
\begin{multline}
\label{eq:Calpha}
    C_{ij}^{(\alpha)}(\tau)= \frac{\hbar^2}{\mu_i \mu_j} \int_0^{\infty} \mathrm{d} \omega [(1+n_\omega^{(\alpha)}) \mathcal{J}_{ij}^{(\alpha)}(\omega)e^{-i\omega \tau}+\\
    n_\omega ^{(\alpha)}\mathcal{J}_{ij}^{(\alpha)*}(\omega)e^{+i\omega \tau}],
\end{multline}
where $n_\omega^{(\alpha)}=({e^{\beta_{\alpha}\hbar\omega}-1})^{-1}$,
and $\beta_{\alpha}=1/(k_BT_0^{(\alpha)})$. It is convenient to rewrite \ref{eq:Calpha} as
\begin{equation}
C^{(\alpha)}_{ij}(\tau)= \frac{\hbar^2}{\mu_i \mu_j} \int_{
-\infty}^{+\infty} \mathrm{d} \omega \,\mathrm{J}^{(\alpha)}_{ij}(\omega; \beta_\alpha)e^{-i\omega \tau},
\end{equation}
where $\mathrm{J}^{(\alpha)}_{ij}(\omega;\beta_\alpha)$, defined for $-\infty<\omega<+\infty$, is
\begin{equation}
\label{eq:Jeff}
     \quad \; \mathrm{J}^{(\alpha)}_{ij}(\omega;\beta_\alpha) = 
\begin{cases}
			(1+n_\omega^{(\alpha)}) \mathcal{J}_{ij}^{(\alpha)}(\omega) & \text{for}\, \omega \geq 0,\\
            n_{|\omega|}^{(\alpha)} \mathcal{J}_{ij}^{(\alpha)*}(|\omega|) & \text{for}\, \omega \leq 0.
		 \end{cases}
 \end{equation}
The function $\mathrm{J}^{(\alpha)}_{ij}(\omega;\beta_\alpha)$ is continuous at $\omega=0$ because the spectral density matrices go to zero for $\omega \rightarrow0$ at least as $\omega^2$. Consequently, the correlation function $C_{ij}(\tau)$
can be expressed as
\begin{equation}
   C_{ij}(\tau)=\frac{\hbar^2}{\mu_i \mu_j} \int_{-\infty}^{+\infty} d\omega \mathrm{J}^\text{eff}_{ij}(\omega;\beta_M , \beta_S) e^{-i\omega \tau},
\end{equation}
where
\begin{equation}
\label{eq:eff_1}
   \mathrm{J}^\text{eff}_{ij}(\omega;\beta_M, \beta_S)= \mathrm{J}^{(M)}_{ij}(\omega;\beta_M)+\mathrm{J}^{(S)}_{ij}(\omega;\beta_S).
\end{equation}

When the medium-assisted reservoir and the scattering-assisted reservoir are at the same temperature $T_0$ we obtain
\begin{equation}
\label{eq:eff_2}
     \quad \; \mathrm{J}^\text{eff}_{ij}(\omega;\beta) = 
\begin{cases}
			\frac{1}{2\pi}(1+n_\omega) \Gamma_{ij}(\omega) & \text{for}\, \omega \geq 0,\\
            \frac{1}{2\pi}n_{|\omega|}\Gamma_{ij}^{*}(|\omega|) & \text{for}\, \omega \leq 0,
		 \end{cases}
 \end{equation}
where now $ n_\omega=(e^{\beta\hbar\omega}-1)^{-1},$ and $\beta=1/(k_BT_0)$. In the zero-temperature limit, expression \ref{eq:eff_2} reduces to
\begin{equation}
\label{eq:eff_3}
     \quad \; \mathrm{J}^\text{eff}_{ij}(\omega;\beta \rightarrow \infty) = 
\begin{cases}
			\frac{1}{2\pi}\Gamma_{ij}(\omega) & \text{for}\, \omega \geq 0,\\
            0 & \text{for}\, \omega \leq 0.
		 \end{cases}
 \end{equation}

\subsection{Surrogate Environment Initially in Vacuum Quantum State}
Let $\mathrm{J}^{(M)}(\omega; \beta_M)$ and $\mathrm{J}^{(S)}(\omega; \beta_S)$ denote the $N\times N$ complex matrices with elements $\mathrm{J}^{(M)}_{ij}(\omega; \beta_M)$ and $\mathrm{J}^{(S)}_{ij}(\omega; \beta_S)$, respectively. We call these matrices \textit{temperature-dependent spectral density matrices} of the medium and scattering-assisted reservoirs, respectively. These matrices are defined for $-\infty<\omega<+\infty$, they are Hermitian and positive definite. We now introduce the \textit{effective spectral density matrix}
\begin{equation}
\label{eq:Sdtd}
\mathrm{J}^\text{eff}(\omega; \beta_M, \beta_S)=\mathrm{J}^{(M)}(\omega; \beta_M)+\mathrm{J}^{(S)}(\omega; \beta_S),
\end{equation}
which characterizes the overall electromagnetic environment. This matrix is also defined for $-\infty<\omega<+\infty$, Hermitian, and positive definite.

By the equivalence established above, the reduced dynamics of the emitters can be computed by considering the surrogate Hamiltonian
\begin{equation}
\label{eq:sur}
    \hat{H}^\text{sur}= \hat{H}_A +  \hat{H}_E^\text{sur}+ \hat{H}_I^\text{sur},
\end{equation}
where $\hat{H}_E^\text{sur}$ is the bare Hamiltonian of a surrogate environment consisting of $N$ bosonic reservoirs, with positive and negative frequencies, initially in the vacuum state,
\begin{equation}
\label{eq:Hem_eq}
\hat{H}_E^\text{sur}=\sum_{j=1}^N \int_{-\infty}^{+\infty} d\omega \hbar\omega \hat{a}_j^\dagger (\omega) \hat{a}_j(\omega);
\end{equation}
here $\hat{a}_j(\omega)$ and $\hat{a}_j^\dagger(\omega)$, with $j=1,\ldots,N$, are the annihilation and creation operators of the bosonic modes of the surrogate environment. The interaction Hamiltonian $\hat{H}_I^\text{sur}$ between the quantum emitters and the surrogate environment is given by
\begin{equation}
\label{eq:Hint_eq}
\hat{H}_I^\text{sur}=-\sum_{i,j=1}^N\hat{\mu}_i  \int_{-\infty}^{+\infty} d\omega [G^\text{eff}_{ij}(\omega)\hat{a}_j(\omega)+h.c.],
\end{equation}
where $G^\text{eff}_{ij}(\omega)$ is the $ij$-th element of the $N\times N$ coupling matrix $G^\text{eff}(\omega)$ given by
\begin{equation}
\label{eq:Gmat}
     {G}^\text{eff}(\omega)= \hbar [P^{-1} \mathrm{J}^\text{eff}(\omega;\beta_M,\beta_S)P^{-1}]^{1/2}.
\end{equation}

When the medium-assisted reservoir and the scattering-assisted reservoir are initially in the vacuum state, the effective spectral density $\mathrm{J}_{ij}^\text{eff}(\omega)$ is given by \ref{eq:eff_3}. In the literature on multiple quantum emitters interacting with dielectric objects based on the macroscopic quantum electrodynamic approach using the Langevin noise formalism, this expression is  widely used (e.g. \cite{sanchez-barquilla_few-mode_2022}, \cite{miguel-torcal_photon-mediated_2025}). However, as already pointed out in \cite{miano_quantum_2025} and \cite{miano_spectral_2025} for a single quantum emitter, the conditions under which \ref{eq:eff_3} remains valid are not always clearly stated. In fact, when the two reservoirs are initially in thermal quantum states at non-zero temperature, the expression \ref{eq:eff_3} is no longer valid. If the temperatures of the two reservoirs are the same, the elements of the effective spectral density matrix are given by \ref{eq:eff_2}. When the initial temperatures are different, the elements of the effective spectral density matrix are given by \ref{eq:eff_1}.

\begin{figure}
    \centering    \includegraphics[width=\linewidth]{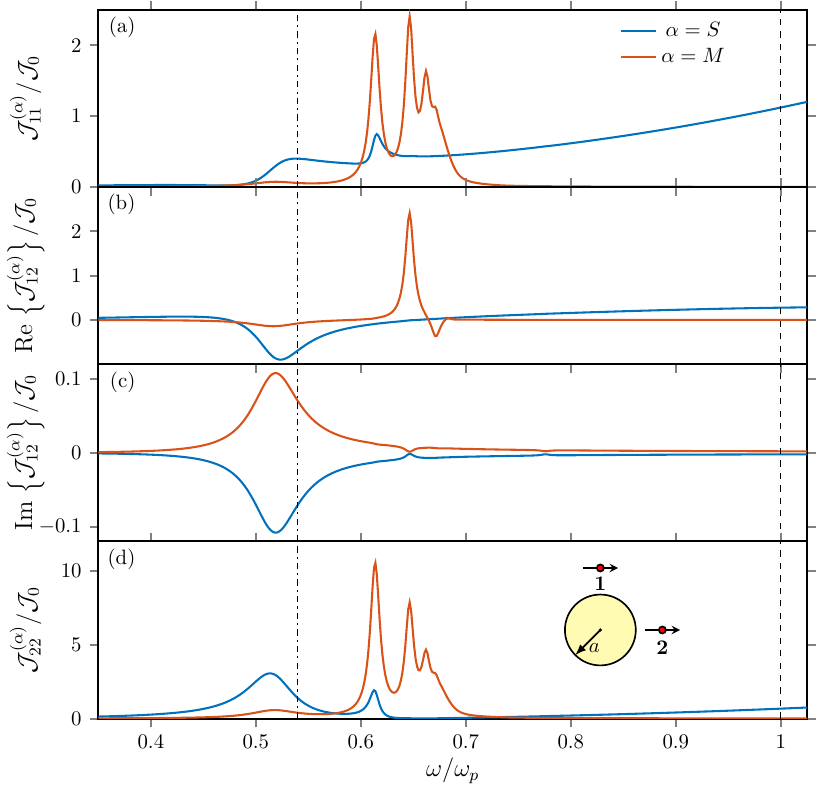}
\caption{Two equal quantum emitters with dipole moment $\mu$ interact with a Drude sphere with radius $a$, plasma frequency $\omega_p$ and damping rate $\nu$. Normalized spectral densities $\mathcal{J}^{(M)}_{ij}/\mathcal{J}_0$ and $\mathcal{J}^{(S)}_{ij}/\mathcal{J}_0$, for $i,j=1,2$, as a function of the normalized frequency $\omega/\omega_p$ with $k_p \, a=1$, $\nu/\omega_p = 0.01$ where $k_p=\omega_p/c$. The two emitters are positioned as in the inset, at the same distance $d$ from the sphere center,
    $d/a=1.75$. The characteristic spectral density $\mathcal{J}_0$ is given by $\mathcal{J}_0 = \frac{1}{6 \pi^2} \frac{k_p}{\hbar\, \varepsilon_0 } \left({\mu k_p}\right)^2$. The spectral density matrices are Hermitian.} 
    \label{fig:Sphere}
\end{figure}

\begin{figure}
    \centering    \includegraphics[width=\linewidth]{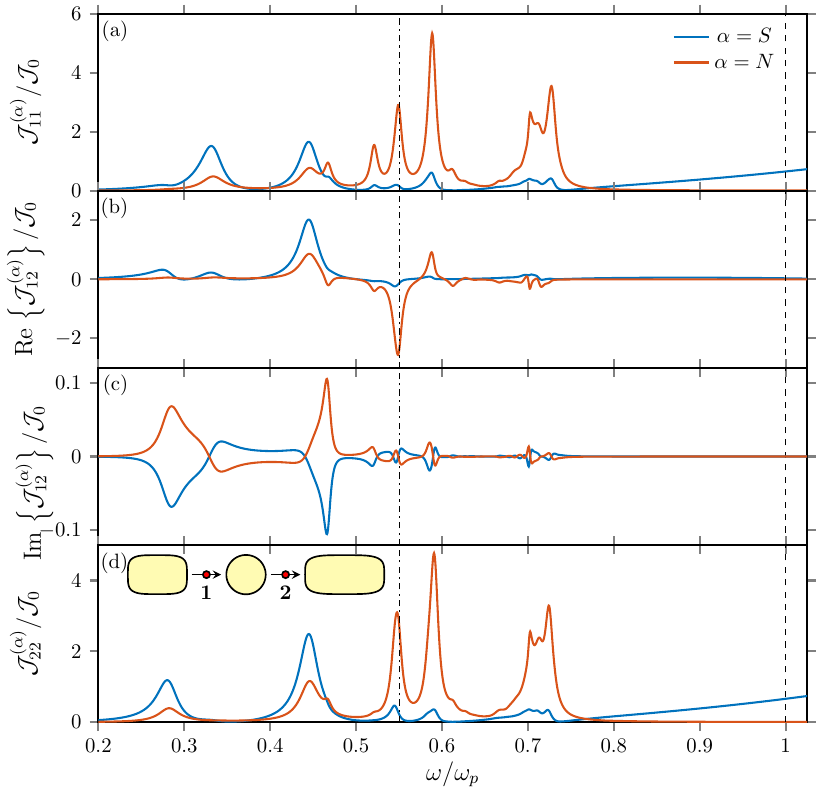}
    \caption{ Two equal quantum emitters with dipole moment $\mu$ interact with a nanostructure composed of two rods and a disk. The first rod has length $3a$, width $2a$, and height $a$;
    the second rod has length $4a$, width $2 a$, and height $a$; the disk has radius $a$ and height $a$. All three particles are made of a Drude material with 
    plasma frequency $\omega_p$ and damping rate $\nu$ such as $\nu/\omega_p = 0.01$. The two quantum emitters are placed at the midpoint of the gaps between adjacent particles.
    Normalized spectral densities $\mathcal{J}^{(S)}_{ij}/\mathcal{J}_0$ and $\mathcal{J}^{(M)}_{ij}/\mathcal{J}_0$ versus $\omega/\omega_p$ for $i,j=1,2$ and $k_p \, a=1$ where $k_p=\omega_p/c$. The characteristic spectral density $\mathcal{J}_0$ is given by $\mathcal{J}_0 = \frac{1}{6 \pi^2} \frac{k_p}{\hbar\, \varepsilon_0 } \left({\mu k_p}\right)^2$. The spectral density matrices are Hermitian. }   
    \label{fig:Rod}
\end{figure}

\section{Application to a system with two quantum emitters}
\label{sec:Application}

We illustrate the impact of a structured electromagnetic environment, formed by dispersive dielectric particles embedded in vacuum, on the dynamics of an emitter ensemble by focusing on the simplest nontrivial case: two quantum emitters, labeled 1 and 2. We  consider two representative nanophotonics settings: a spherical particle in vacuum, as in \cite{miano_spectral_2025}, and a nanostructure composed of two rods and a disk in vacuum (see the insets of Figs. \ref{fig:Sphere} and \ref{fig:Rod}, respectively). The two quantum emitters are placed at the positions indicated by the red dots in the insets. All particles are homogeneous and their dielectric permittivity is described by the Drude model: $\varepsilon_\omega=[1-\omega_p^2/(\omega^2+i\nu\omega)]$ with plasma frequency  $\omega_p$  and relaxation frequency $\nu$.

Within the modified Langevin noise formalism and the emitter-centered description, the electromagnetic environment is modeled by two continuous bosonic baths, medium-assisted  (M) and scattering-assisted  (S) reservoirs, characterized by the spectral density matrices $\mathcal{J}^{(M)}(\omega)$ and $\mathcal{J}^{(S)}(\omega)$. The two reservoirs are initially in thermal quantum states at inverse temperatures $\beta_{\text{M}}$ and $\beta_\text{S}$. The two emitters interact through their dipole moments with the continuous sets of bosonic excitations of the electromagnetic environment. 

Each emitter is modeled as a two-level system with transition frequency $\Omega_i$ and dipole moment magnitude $\mu_i$ ($i=1,2$). Hence, the bare quantum emitter Hamiltonian $\hat{H}_A$ is given by
\begin{equation}
\label{eq:Hs}
   \hat{H}_A=\frac{\hbar \Omega_1}{2}\hat{\sigma}_z^{(1)}+ \frac{\hbar \Omega_2}{2}\hat{\sigma}_z^{(2)},
\end{equation}

and the transition dipole-moment operators are $\hat{\mu}_1=\mu_1 \hat{\sigma}^{(1)}_x$ and $\hat{\mu}_2=\mu_2 \hat{\sigma}^{(2)}_x$. Here $\hat{\sigma}_z^{(i)}=\ket{e^{(i)}}\!\bra{e^{(i)}}-\ket{g^{(i)}}\!\bra{g^{(i)}}$,
$\hat{\sigma}_x^{(i)}=\ket{g^{(i)}}\!\bra{e^{(i)}}+\ket{e^{(i)}}\!\bra{g^{(i)}}$
where $\ket{g^{(i)}}$ and $\ket{e^{(i)}}$ denote, respectively, the ground and excited states of the $i$-th quantum emitter. In the numerical examples below, we considered equal dipole moments  $\mu_1=\mu_2\equiv \mu$.

As shown in Sec. \ref{sec:Correlator}, leveraging the temperature-dependent spectral densities $\mathrm{J}^{(M)}(\omega; \beta_M)$ and $\mathrm{J}^{(S)}(\omega; \beta_S)$, the degrees of freedom of the medium-assisted and scattering-assisted baths are reduced to those of two distinct sets of surrogate continuous bosonic reservoirs initially in the vacuum quantum state and with an effective spectral density matrix $\mathrm{J}^\text{eff}(\omega; \beta_M, \beta_S)$. Furthermore, each emitter is coupled to both continuous reservoirs through direct coupling $G^{\text{eff}}_{ii}(\omega)$ and cross-coupling amplitudes $G^{\text{eff}}_{ij}(\omega)$. The coupling of two emitters with a common reservoir has been known to generate quantum correlated states of the composite system \cite{braun_2002}, similar to what has been studied in the field of cQED \cite{Blais_2004}.  More generally, for linear
electromagnetic environments the effective description
involves a number of reservoirs equal to the number of emitters, and each emitter is simultaneously coupled to all of them.
We study the non-equilibrium dynamics of the emitters for a set of physically relevant initial  quantum states, including pure
product states and maximally entangled Bell states.

We investigate the reduced dynamics of the two quantum emitters using the Matrix Product State approach, similarly to what is done with the TEDOPA method \cite{tamascelli_efficient_2019},
to simulate the time evolution of the whole emitter + surrogate environment state governed by the effective Hamiltonian defined by Eqs. \ref{eq:sur}-\ref{eq:Hs}. Such methods have been shown to reproduce exact results and to provide corrections to more recent numerical techniques based on collisional models \cite{Lacroix_2025}.

\subsection{Spectral Densities}
 
We begin by analyzing the behavior of the normalized spectral densities $\mathcal{J}_{ij}^{(M)}(\omega)/\mathcal{J}_0$ and $\mathcal{J}_{ij}^{(S)}(\omega)/\mathcal{J}_0$ as a function of $\omega/\omega_p$ for $i,j\in\{1,2\}$, where
\begin{equation}
\mathcal{J}_0 = \frac{1}{6\pi^2} \frac{k_p}{\hbar\, \varepsilon_0} \left({\mu k_p}\right)^2,
 \end{equation}
 and $k_p=\omega_p/c$. By definition, the characteristic spectral density $\mathcal{J}_0$ is equal to the value of the vacuum spectral density at the wavenumber $k_p$, and indicates the coupling strength between the quantum emitter and the vacuum field. Since the spectral density matrices are Hermitian, we omit the plots of $\mathcal{J}_{21}^{(M)}(\omega)$ and $\mathcal{J}_{21}^{(S)}(\omega)$ .
We evaluated numerically the spectral density matrices by using a surface integral equation formulation, as described in Appendix \ref{sec:SIE}.

Figure \ref{fig:Sphere} shows $\mathcal{J}^{(S)}_{ij}/\mathcal{J}_0$ and $\mathcal{J}^{(M)}_{ij}/\mathcal{J}_0$ for a Drude sphere with radius $a$ as a function of the normalized frequency $\omega/\omega_p$, assuming size parameter $k_p \, a=1$ and normalized damping rate $\nu/\omega_p = 0.01$. The two emitters are positioned as in the inset, each at a distance $d$ from the center of the sphere, with $d/a=1.75$. In the low–frequency region around the dipolar plasmon resonance ($\omega/\omega_p\simeq 0.54$), the amplitude of the elements $\mathcal{J}^{(S)}_{ij}$ is comparable or larger than $\mathcal{J}^{(M)}_{ij}$. At higher frequencies, near higher-order plasmonic resonances, the amplitude of $\mathcal{J}^{(M)}_{ij}$ becomes significantly larger than that of $\mathcal{J}^{(S)}_{ij}$. For a comprehensive analysis of the role of the scattering modes in shaping the overall spectral density, see Ref. \cite{miano_spectral_2025}. The bandwidths of the plasmonic modes are limited by a combination of radiative and material losses, as detailed in \cite{miano_spectral_2025}. 

Figure \ref{fig:Rod} shows the corresponding normalized spectral densities for a nanostructure composed of two rods and a disk. The first rod has length $3a$, width $2 a$, and height $a$; the second rod has length $4a$, width $2 a$, and height $a$; the disk has radius $a$ and height $a$. The emitters are placed at the midpoint of the gaps between adjacent particles. All materials follow the same Drude model with $k_p \, a=1$ and a damping rate  $\nu/\omega_p = 0.01$. Also in this case, the scattering-assisted spectral densities are typically larger than medium-assisted at low frequencies, whereas at higher frequencies the medium-assisted contribution dominates and is typically characterized by sharper peaks associated with higher multipolar scattering orders.

The effective spectral density matrix $\mathrm{J}^\text{eff}(\omega; \beta_M, \beta_S)$ completely characterizes the interaction of the two emitters with the surrogate bosonic environment through Eq. \ref{eq:Gmat}: it is the sum of temperature-dependent medium-assisted $\mathrm{J}^{(M)}(\omega; \beta_M)$ and temperature-dependent scattering-assisted $\mathrm{J}^{(S)}(\omega; \beta_S)$ spectral density matrices. They are proportional to the medium- and scattering-assisted spectral density matrices and depend on the temperatures through the Bose-Einstein distribution according to Eq. \ref{eq:eff_2} .

\subsection{Entanglement Decay and Revivals}\label{subsec:revivals}

Since the early days of quantum information theory \cite{Nielsen_Chuang_2010}, quantum entanglement~\cite{Plenio_2007,Horodecki2009} has been recognized as a key resource for a variety of tasks \cite{Chitambar_2019}. During the past two decades, its generation \cite{braun_2002,Benatti_2003,Maniscalco_2008,Zell_2009,Ma_2012}, sudden death \cite{Yu_Eberly_2004, YuEberly_2006,YuEberly_2009,Mazzola_2009}, birth \cite{Ficek_2008}, and degradation \cite{Zyczkowski_2001,    Bellomo_degradation2010,Aolita_2015,Cerrato_2025} under multiple external environments \cite{Cattaneo_2019} have been extensively investigated. In particular, if the quantum bipartite system undergoes non-Markovian dynamics \cite{BellomoNM_2007,Bellomodyn_2008,Bellomotrap_2008,Rivas_2010,Maniscalco_2008,Ma_2012}, it has been shown that the quantum state experiences revivals in its entanglement features. More recently, the generation of entangled states of pairs of quantum 
emitters using plasmonic  \cite{bellomo_creation_2013,gonzalez_tudela_2011}, as well as  photonic 1D waveguides \cite{sheremet_waveguide_2023} through bound states in the continuum \cite{Facchi_2016, Magnifico_2025}, has been studied.

Here, we investigate relaxation and the occurrence of entanglement revivals in the nanophotonic environments described above.
We assume that the two emitters are prepared in the pure, maximally entangled Bell state $\ket{\Psi^{\pm}}=\frac{1}{\sqrt{2}}(\ket{eg}\pm\ket{ge})$, while the surrogate bosonic environment is initially in the vacuum state. 
As the system evolves under the Hamiltonian in \ref{eq:sur}, each emitter undergoes Rabi evolution in the presence of a continuous structured photonic environment \cite{frisk_kockum_ultrastrong_2019}.  Crucially, besides direct coupling to emitter-centered modes, cross-coupling interaction terms are present that mediate environment-induced correlations between the emitters.

We denote the reduced density operator of the two emitters by $\hat{\rho}_{12}$. We quantify entanglement via negativity $\mathcal{N}(\hat{\rho}_{12})=( \|\hat{\rho}_{12}^{T_1}\|_{1}-1)/2$, where $\hat{\rho}_{12}^{T_1}$ is the partial transpose of $\hat{\rho}_{\text{12}}$ with respect to quantum emitter 1 and ${\| \hat{O}\|}_{1}=\tr\sqrt{\hat{O}^{\dagger}\hat{O}}$ is the trace norm. Negativity is a widely used measure of the inseparability of bipartite quantum states \cite{Vidal_2002}. Since the local Hilbert space dimension of both emitters is two, the positivity of the partial transpose, i.e., the PPT criterium, is necessary and sufficient for separability \cite{Peres_1996,HORODECKI19961}. 

\begin{figure}
    \centering    
    \includegraphics[width=\linewidth]{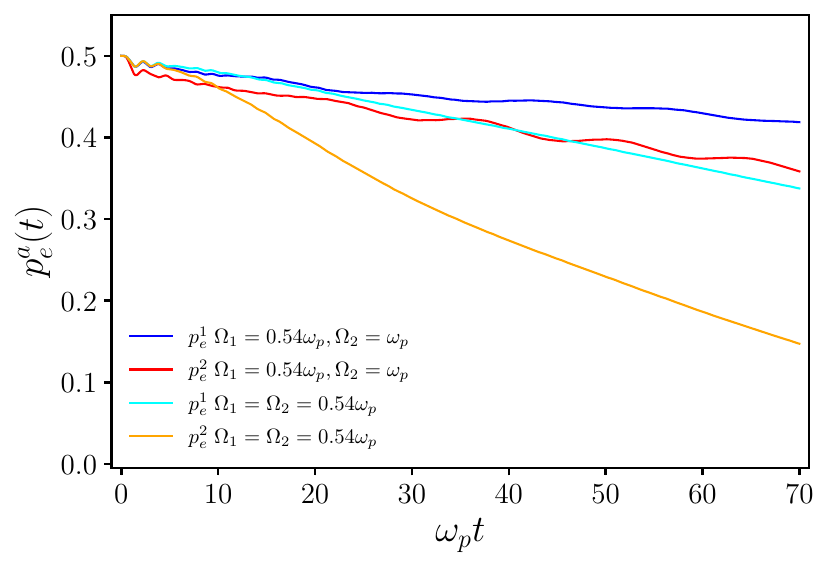}
    \caption{Spherical particle, see Fig. \ref{fig:Sphere}. Relaxation dynamics of the excited-state populations $p_e^{a}(t)$ for the two quantum emitters ($a=1,2$). The initial composite state is $\hat{\rho}_{12}(0)=\ketbra{\Psi^{-}}$, the parameters are $\mu=10^{-3}\omega_{p}$ and $\beta_{S}=\beta_{M}=1000$. Blue (red) curves refer to emitter $a=1$ ($a=2$) for $\Omega_1=0.54\omega_p,\Omega_2=\omega_p$, while cyan (orange) curves refer to emitter $a=1$ ($a=2$) for $\Omega_1=\Omega_2=0.54\omega_p$ for emitter $a=1$ ($a=2$). The cutoff frequency of the environment is set to  $\omega_{\text{cut}}=4.0 \, \omega_p$.}   
    \label{fig:popAB1}
\end{figure}

\begin{figure}
    \centering    \includegraphics[width=\linewidth]{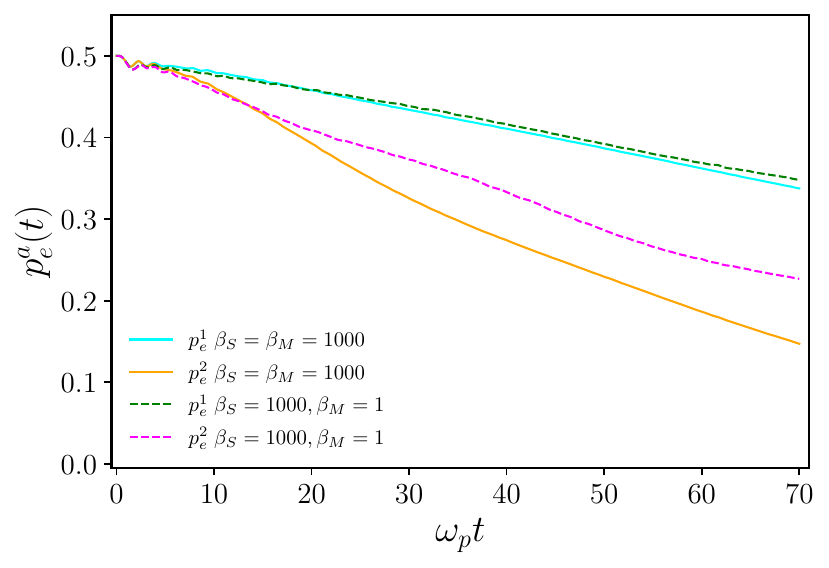}
    \caption{Spherical particle, see Fig. \ref{fig:Sphere}. Relaxation dynamics of the excited-state populations $p_e^{a}(t)$ for the two quantum emitters ($a=1,2$). The initial composite state is set to $\rho_{12}(0)=\ketbra{\Psi^{-}}$, while $\mu=10^{-3}\omega_{p}$ and $\Omega_1=\Omega_2=0.54\omega_p$. Solid curves are correspond to $\beta_{S}=\beta_{M}=1000$, whereas dashed curves illustrate the case $\beta_{M}=1,\beta_{S}=1000$.  The cutoff frequency of the environment is set to $\omega_{\text{cut}}=4.0\omega_p$.}   
    \label{fig:popAB2}
\end{figure}

We examine the time evolution of the reduced states of the emitters and of the negativity in the course of relaxation. We first focus on the dielectric sphere in Fig. \ref{fig:Sphere}. The transition frequencies are chosen either resonant, $\Omega_{1}=\Omega_{2}=0.54 \omega_{p}$ (dot-dashed vertical line in Fig. \ref{fig:Sphere}), or detuned, with $\Omega_{1}=0.54\,\omega_{p}$ and $\Omega_{2}=\omega_{p}$ (dashed vertical line). In Figs. \ref{fig:popAB1} and \ref{fig:popAB2}, we show the dynamics of the reduced density operators $\hat{\rho}_{\text{1(2)}}(t)=\Tr_{2(1)}[\hat{\rho}_{\text{12}}(t)]$, focusing on the populations of the excited states, $p^{a}_{e}(t)=\langle e|\hat{\rho}_a(t)|e\rangle$ where $a=1,2$. For weak coupling and low temperatures of medium and scattering-assisted reservoirs,  each reduced state  evolves from totally mixed states to mixed states with $p_g^a>p_e^a$ where $p^{a}_{g}(t)=\langle g|\hat{\rho}_a(t)|g\rangle$. After a transient marked by nonmonotonic behavior, which is typical of non-Markovian relaxation \cite{bassano_vacchini_open_2024}, each emitter decays with different rates, which depend on the physical features of the spectral densities of the environment. The relaxation rate of the reduced state strongly depends on the detuning of the transition frequencies: when $\Omega_1=\Omega_2$, the emission is enhanced due to the degeneracy of the eigenstates of $H_{\text{A}}$, i.e. $\ket{g,e},\ket{e,g}$. It can be traced back to the combined effect of first- and second-order processes on the coupling strength. Second-order effects are marked by virtual transitions made of emission and reabsorption of photonic excitations. For near-resonant emitter frequencies, they can become comparable to first-order emission processes.

Allowing medium- and scattering-assisted reservoirs to be initially in thermal states at different temperatures, with the medium-assisted reservoir playing the role of the hot bath (Fig. \ref{fig:popAB2}) yields larger excited-state populations for both emitters.  This behavior is explained by the increased probability of photon absorption from the warmer reservoir, which partially counteracts emission.

\begin{figure}
    \centering    \includegraphics[width=\linewidth]{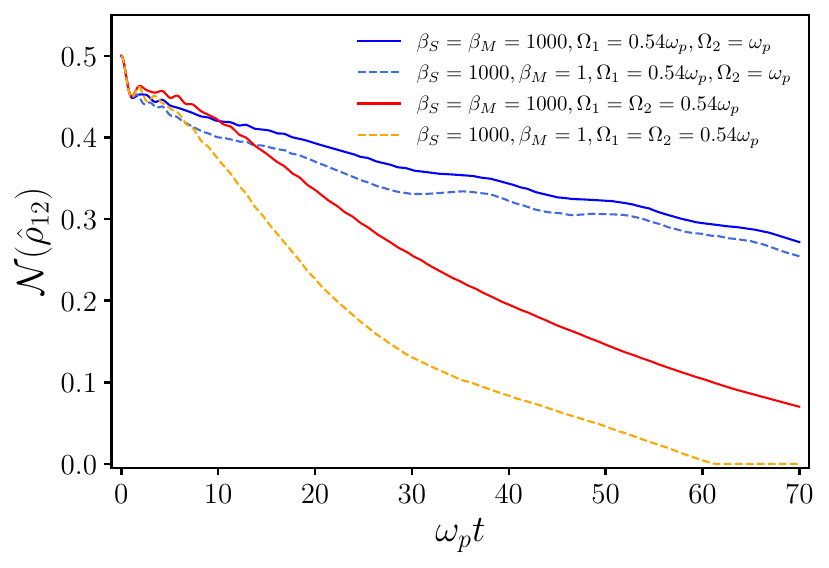}
    \caption{Spherical particle, see Fig. \ref{fig:Sphere}. Time evolution of negativity $\mathcal{N}(\hat{\rho}_{12})$ with $\mu=10^{-3}\omega_{p}$, corresponding to the same initial state as in Figs. \ref{fig:popAB1} and \ref{fig:popAB2}. Solid lines denote the cases $\Omega_1=0.54\omega_p,\Omega_2=\omega_p$ and $\Omega_1=\Omega_2=0.54\omega_p$ with equal inverse temperatures $\beta_{S}=\beta_{M}=1000$, respectively, while dashed curves denote the same frequency choices for $\beta_{M}=1$ and $\beta_{S}=1000$. The cutoff frequency of the environment is set to  $\omega_{\text{cut}}=4.0\omega_p$.} 
    \label{fig:negtysphere}
\end{figure}

The reduced-state dynamics in Figs. \ref{fig:popAB1} and \ref{fig:popAB2} stem from the evolution of the initially pure entangled state of the composite system  under coupling to the photon reservoirs.  As the system relaxes, the state becomes progressively mixed, while initial quantum correlations, such as entanglement, are progressively degraded. This process is illustrated in Fig. \ref{fig:negtysphere}, where the entanglement negativity $\mathcal{N}(\hat{\rho}_{12})$ is plotted as a function of time for different temperatures of the thermal states of the electromagnetic environment and for different detuning of the emitter transition frequencies. The detrimental impact of finite-temperature bath states on negativity is due to the combined effect of increased populations of excited emitter states and of the enhanced decay of two-body correlations, sustained by off-diagonal terms of $\hat{\rho}_{12}$. The Rabi-like interaction with the electromagnetic environment thus leads the composite state to a separable state $\mathcal{N}(\hat{\rho}_{12})=0$. This relaxation is governed by the interplay between direct- and cross-correlations, both of which extend over a broad frequency range due to the presence of the scattering-assisted reservoir.

\begin{figure}
    \centering    
    \includegraphics[width=\linewidth]{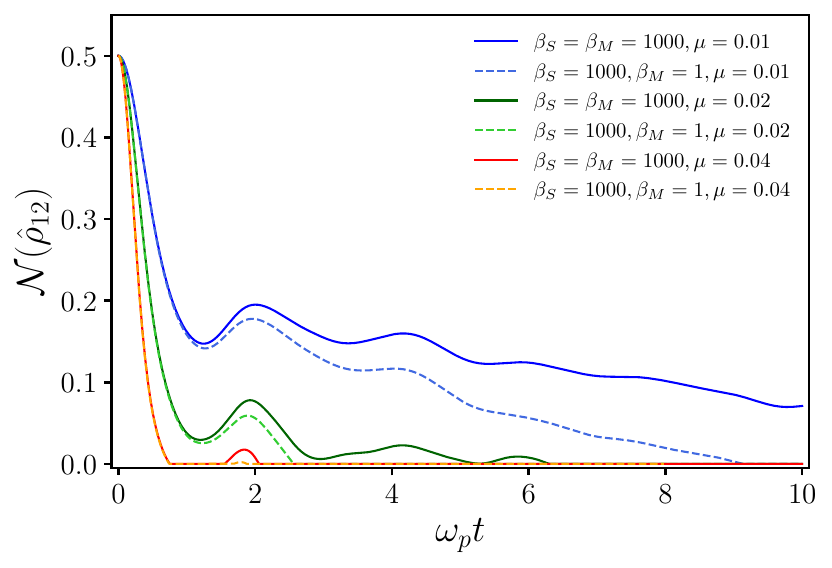}
    \caption{Spherical particle, see Fig. \ref{fig:Sphere}. Time evolution of negativity $\mathcal{N}(\hat{\rho}_{12})$, computed for $\Omega_1=0.54\omega_p,\Omega_2=\omega_p$ and the same initial state as in Figs.  \ref{fig:popAB1} and \ref{fig:popAB2}. Solid lines denote the cases $\mu=0.01,0.02,0.04$ and inverse temperatures $\beta_{S}=\beta_{M}=1000$, while dashed lines denote analogous curves computed for $\beta_{M}=1,\beta_{S}=1000$. The cutoff frequency of the environment is set to  $\omega_{\text{cut}}=4.0\omega_p$.} \label{fig:negtyspherestrong1}
\end{figure}

With stronger coupling, the dynamics becomes richer. Figure \ref{fig:negtyspherestrong1} shows the behavior of $\mathcal{N}(\hat{\rho}_{12})$ for different values of the dipole momentum $\mu$, inverse temperatures, and fixed detuning of the emitter frequencies. Negativity undergoes sudden death at short times, followed by revivals of decreasing amplitudes at later times, a hallmark of non-Markovian effects \cite{Ma_2012}, arising from emitter-photon correlations, which are present even in the RWA limit \cite{BellomoNM_2007}. 
In the present model, counter-rotating terms in Eq.~\eqref{eq:Hint_eq} tend to suppress revival amplitudes compared to a purely RWA description. In addition, the finite-temperature states of the medium-assisted reservoir noticeably curb the entanglement revivals. 

\begin{figure}
    \centering    
    \includegraphics[width=\linewidth]{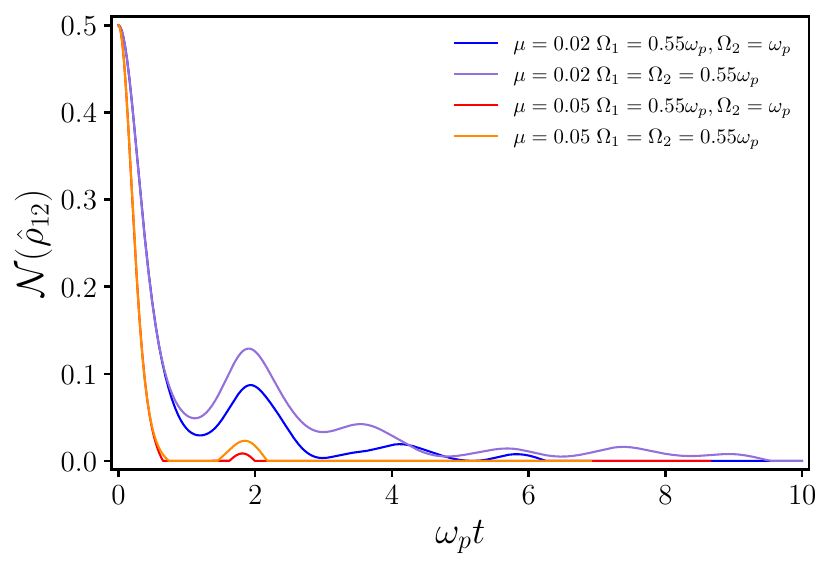}
    \caption{Rod-nanodisk configuration, see Fig. \ref{fig:Rod}. Time evolution of negativity $\mathcal{N}(\hat{\rho}_{12})$, computed for $\beta_M=\beta_S=1000$ and the same initial state as in Figs.\ref{fig:popAB1} and \ref{fig:popAB2}, for different choices of  transition frequencies and dipole momenta. The cutoff frequency of the environment is set to  $\omega_{\text{cut}}=4.0\omega_p$.} \label{fig:negtyrod}
\end{figure}

 The analysis of negativity is repeated in Fig. \ref{fig:negtyrod} for the configuration of Fig. \ref{fig:Rod}, where two emitters interact with two rods and a nano-disk, at equal, low temperatures of the medium and scattering-assisted reservoirs. Transient, short time entanglement revivals exhibit a marked dependence on detuning:  for zero detuning and moderate coupling, the revivals are enhanced at short times. Altogether, these results highlight how the detailed structure of the medium- and scattering-assisted spectral density matrices affects the degradation and revival of entanglement, suggesting concrete routes to optimize quantum correlations in structured nanophotonic environments.

\subsection{Entanglement Generation}
\label{subsec:gener} 
We now address the creation of entanglement starting from separable state preparations. In particular, we consider initial two-emitter states that are convex mixtures of product states $\sum_{k} p_{k}\,\hat{\rho}^{k}_{1} \otimes \hat{\rho}^{k}_{2}$, and examine whether entanglement emerges in the course of relaxation, keeping in mind that separable states may still display quantum correlations beyond entanglement \cite{Adesso_2016}. To isolate environment-induced effects, we take the initial two-emitter state to be the pure product $\hat{\rho}_{12}(0)=\ketbra{eg}$, so that no initial classical or quantum correlations are present. The electromagnetic environment is associated with the dielectric sphere of Fig. \ref{fig:Sphere}; where we set the detuning of the two emitter frequencies to zero, i.e., $\Omega_1=\Omega_2=0.54 \omega_{p}$.

\begin{figure}
    \centering    
    \includegraphics[width=\linewidth]{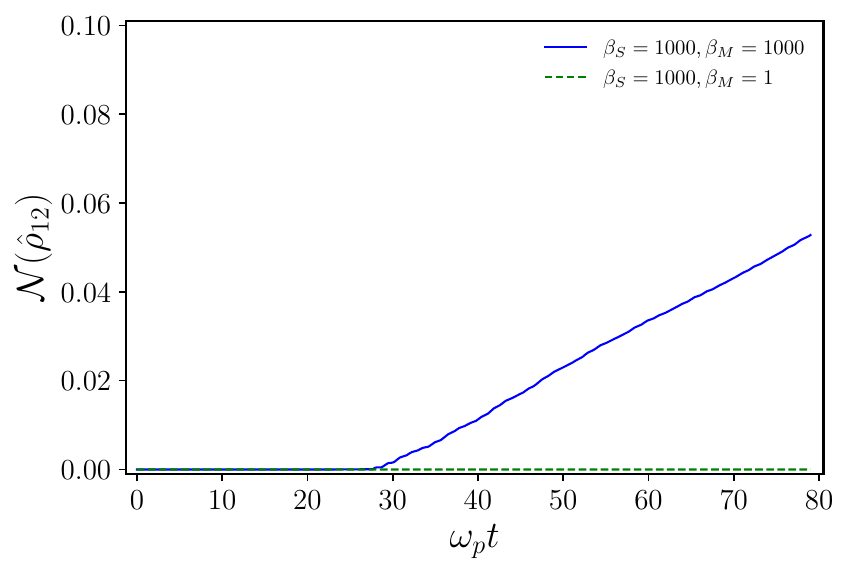}
    \caption{Spherical particle, see Fig. \ref{fig:Sphere}. Time evolution of negativity $\mathcal{N}(\hat{\rho}_{12})$ for the initial composite state $\hat{\rho}_{12}(0)=\ketbra{eg}$, $\Omega_1=\Omega_2=0.54\omega_p$, $\mu=10^{-3}\omega_p$, computed for $\beta_{S}=\beta_{M}=1000$ (solid line) and  $\beta_{M}=1,\beta_{S}=1000$ (dashed line). The cutoff frequency of the environment is set to  $\omega_{\text{cut}}=4.0\omega_p$.}    
    \label{fig:negty01beta}
\end{figure}

Figure \ref{fig:negty01beta} shows the evolution of $\mathcal{N}(\hat{\rho}_{12})$ for fixed coupling strengths and inverse temperatures. At sufficiently long times, the negativity $\mathcal{N}(\hat{\rho}_{12})$ becomes positive and then grows, demonstrating that entanglement between the two emitters generated dynamically from an initially separable state. This behavior originates from the cross-coupling terms in Eq. \eqref{eq:Hint_eq}, which allow photon-mediated interactions between the two emitters. Similarly to Sec.\ref{subsec:revivals}, in the case of finite temperature of the medium-assisted reservoir, the system evolves in a separable state for very long times, i.e., in the considered setting, finite temperatures turn out to be detrimental for mediated interactions. 

\begin{figure}
    \centering    \includegraphics[width=\linewidth]{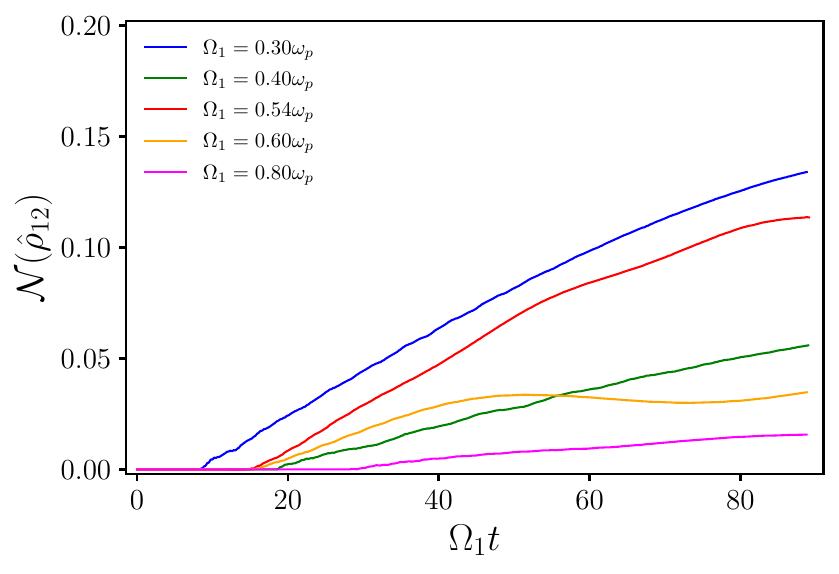}
    \caption{Spherical particle, see Fig. \ref{fig:Sphere}. Time evolution of negativity $\mathcal{N}(\hat{\rho}_{12})$ for the initial composite state $\hat{\rho}_{12}(0)=\ketbra{eg}$, $\mu=10^{-3}\omega_p$, $\beta_{S}=\beta_{M}=1000$, computed for different values of the emitter frequencies $\Omega_1=\Omega_2$ in the range $[0.30\omega_p, 0.80\omega_p]$. The cutoff frequency of the environment is set to  $\omega_{\text{cut}}=4.0\omega_p$.}  
\label{fig:negty01varomeg}
\end{figure}

An additional dynamical signature of the influence of the cross-coupling interactions along with their spectral features is reported in Fig. \ref{fig:negty01varomeg}, where we compare the negativity  $\mathcal{N}(\hat{\rho}_{12})$ for time-evolved states of identical emitters with $\Omega_1=\Omega_2$ for different values of the common transition frequency.  When the transition frequencies are tuned close to the peak of the cross spectral density $\mathcal{J}_{12}(\omega)$, the value of negativity at moderately long times can be enhanced.
 
\begin{figure}
    \centering    \includegraphics[width=\linewidth]{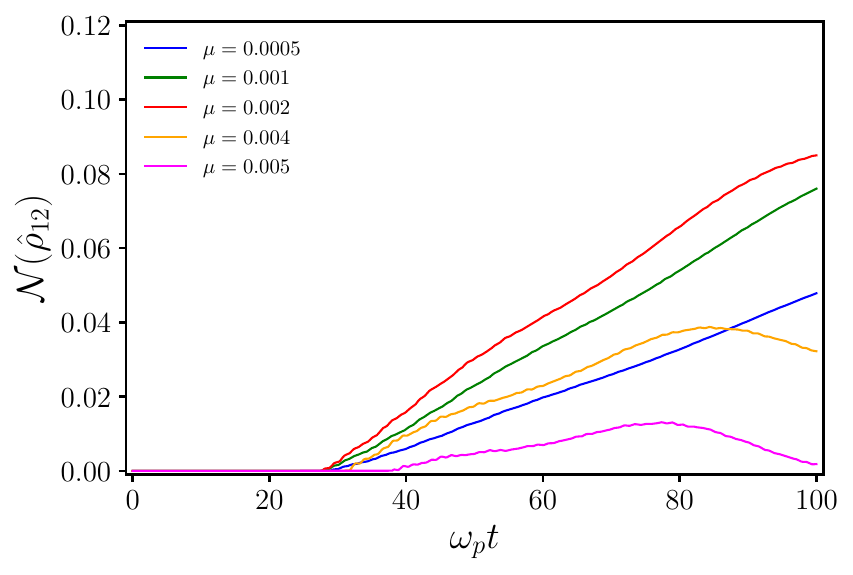}
    \caption{Spherical particle, see Fig. \ref{fig:Sphere}. Time evolution of negativity $\mathcal{N}(\hat{\rho}_{12})$ for the initial composite state $\hat{\rho}_{12}(0)=\ketbra{eg}$, $\Omega_1=\Omega_2=0.54\omega_p$,$\beta_{S}=\beta_{M}=1000$, computed for different dipole strengths $\mu$ in the range $[5\cdot 10^{-4}\omega_p,5\cdot 10^{-3}\omega_p]$. The cutoff frequency of the environment is set to  $\omega_{\text{cut}}=4.0\omega_p$.}   
    \label{fig:negty01varialp}
\end{figure}

It is also natural to ask how the generation of entanglement at long times depends on the coupling strength. 
In principle, increasing the coupling strength of the two emitters can enhance the build-up of long-time quantum correlations. However, due to the form of Eq. \eqref{eq:Hint_eq}, relaxation of the individual emitter states would also increase, as well as decoherence in the state of the composite system. The resulting competition leads to a non-monotonic dependence of the generated entanglement on $\mu$. This behavior is evident in Fig.~\ref{fig:negty01varialp}, where we plot $\mathcal{N}(\hat{\rho}_{12})$ at low environment temperatures for increasing $\mu$. As  $\mu$ grows, the onset of nonzero negativity is delayed. For sufficiently strong coupling, it is evident that the negativity undergoes a transient time behavior: it reaches a local maximum at intermediate times and then decays, approaching zero at long times.
Overall, the data in Fig.~\ref{fig:negty01varialp} point towards the existence of steady states that cannot sustain entanglement between the emitters, at least for moderate coupling strengths.

However, establishing the existence and nature of such steady states would require different numerical approaches \cite{westhoff_2025}, which is left for future work.

\section{Conclusions and outlook}
\label{sec:Conclusion}
Macroscopic quantum electrodynamics (QED) provides an efficient ab-initio framework for describing how quantum emitters interact with structured nanophotonic environments. In this work, we extended the modified Langevin noise approach we recently developed for a single quantum emitter \cite{miano_quantum_2025,miano_spectral_2025} to the case of multiple quantum emitters. Within this formalism, the electromagnetic environment is represented by two continuous bosonic reservoirs, a medium-assisted bosonic reservoir and a scattering-assisted bosonic reservoir, which may, in general, be prepared in different initial quantum states. 

We defined two matrix-valued spectral densities, the medium-assisted and scattering-assisted spectral density matrices, which fully characterize the interaction of the electromagnetic environment with the quantum emitters. Importantly, these matrices cannot be reconstructed from the sole knowledge of the dyadic Green function evaluated at the emitter positions. We showed that they can be computed efficiently within classical electromagnetics by means of surface-integral-equation formulations of linear scattering. 
We then introduced temperature-dependent medium and temperature-dependent scattering-assisted spectral density matrices to describe electromagnetic environments with medium-assisted and scattering-assisted bosonic reservoirs initially in thermal quantum states at different temperatures. Finally, we demonstrated that the time evolution of the reduced density operator of the quantum emitters can be obtained by introducing a surrogate bosonic environment, initially in a vacuum quantum state, whose effective spectral density matrix is the sum of the temperature-dependent medium and temperature-dependent scattering-assisted spectral density matrices. Once this effective spectral density matrix is known, the reduced dynamics of the quantum emitters can be evaluated with standard methods for non-Markovian open quantum systems.

We apply this approach to two quantum emitters coupled to a metallic spherical particle and a metallic nanostructure composed of two rods and a disk. In these representative geometries, we observe entanglement decay and pronounced revivals, as well as the emergence of entanglement from  an initial product state. These findings highlight that the fate of quantum correlations is governed not merely by local decay rates, but by the full, frequency-dependent structure of the environment’s spectral-density matrices, which controls both the degradation and the generation of quantum correlations. More broadly, they indicate that the proposed approach provides a practical route to engineering, and ultimately optimizing, genuinely quantum features of emitter dynamics in structured nanophotonic environments driven out of thermal equilibrium. 

A natural next step is to extend the framework to molecular polaritons, focusing on dilute ensembles of molecular emitters coupled to optical resonators.

\begin{acknowledgments}
This work was supported by the Italian Ministry of University and Research through the PNRR MUR Project No. PE0000023-NQSTI (C.F. and G.M.) and through the PNRR MUR Project No. CN00000013-ICSC (L.M.C.).
\end{acknowledgments}

\appendix

\section{Bright electric field operator}
\label{sec:Bright}
The contribution to the electric field operator due to the bright modes can be expressed in terms of the bosonic operators $\hat{C}_j$ and $\hat{D}_j$. We obtain 
\begin{multline}
    \hat{\mathbf{E}}^\text{bright}(\rb)=\sum_{j=1}^N \int_0^\infty d\omega \left[\mathbf{E}_j^{(M)}(\rb, \omega)\hat{C}_j(\omega)+h.c.\right]+\\
    \sum_{j=1}^N \int_0^\infty d\omega \left[\mathbf{E}_j^{(S)}(\rb, \omega)\hat{D}_j(\omega)+h.c.\right],
\end{multline}
where
\begin{subequations}
\begin{align}
   \mathbf{E}_i^{(M)}(\rb, \omega)&=\sum_{j=1}^N U_{ij}^*(\omega) \mathbf{e}_j^{(M)}(\rb, \omega), \\
   \mathbf{E}_i^{(S)}(\rb, \omega)&=\sum_{j=1}^N V_{ij}^*(\omega) \mathbf{e}_j^{(S)}(\rb, \omega),
\end{align}    
\end{subequations}
and
\begin{subequations}
\begin{align}
   \mathbf{e}_j^{(M)}(\rb, \omega)=\int_V d^3\rb' \mathcal{G}_{e}(\rb,\rb',\omega)\mathcal{G}^{*T}_{e}(\rb',\rb_j,\omega)\mathbf{u}_j, \\
   \mathbf{e}_j^{(S)}(\rb, \omega)= \int do_\mathbf{n} \sum_\nu \mathbf{E}_{\omega \mathbf{n} \nu }(\rb)\mathbf{E}_{\omega \mathbf{n} \nu}^*(\rb_j)\mathbf{u}_j.
\end{align}    
\end{subequations}

\section{Spectral density matrices and power observables in the framework of classical electrodynamics}

We here show that the spectral density matrices, introduced in Section IV, are related to power observables of the system of multiple emitters and dielectric objects in the framework of classical electrodynamics. We consider a reciprocal dielectric object in an unbounded space. The object is driven by $N$ electric dipoles, all oscillating harmonically at the angular frequency $\omega$. The $i$-th electric dipole is located at position $\mathbf{r}_i$ and has dipole moment $\mathbf{p}_i = \mu_i \mathbf{u}_i$ where $\mu_i$, assumed real, is the amplitude and $\mathbf{u}_i$ is a unit vector that gives the dipole orientation. The associated electric current density is $\mathbf{J}_i = - i \omega \mu_i \mathbf{u}_i \, \delta \, (\mathbf{r} - \mathbf{r}_i )$. 
Because the medium is linear, the total electric field generated by
$\mathbf{J} (\mathbf{r}) = \sum_{i=1}^N \mathbf{J}_i  (\mathbf{r})$
is the superposition of the fields produced when each dipole acts alone:
 \begin{equation}
     \mathbf{E} (\mathbf{r}) = \displaystyle \sum_{j=1}^N \mathbf{E}_j (\mathbf{r}).
     \label{eq:EMfield}
 \end{equation}
The electric field $ \mathbf{E}_j$, generated by the $j$-th dipole alone, can be expressed via the dyadic Green's function $\mathcal{G}_\omega$ as
\begin{equation}
    \mathbf{E}_j (\mathbf{r}) = \frac{\omega^2 }{\varepsilon_0 c^2} \, \displaystyle \mathcal{G}_\omega (\mathbf{r}, \mathbf{r}_j) \cdot (\mu_j \mathbf{u}_j). 
    \label{eq:Etot}
\end{equation}

\begin{figure}
    \centering
    \includegraphics[width=0.75\linewidth]{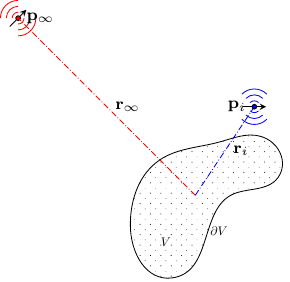}
    \caption{Application of the Lorentz reciprocity theorem to two source-field configurations.  In Scenario~I, only dipole \(\mathbf{p}_i=\mu_i\mathbf{u}_i\) at \(\mathbf{r}_i\) is present, producing fields \((\mathbf{E}_i,\mathbf{H}_i)\). In Scenario~II, a reference dipole \(\mathbf{p}_\infty=\mu_\infty\mathbf{u}_\infty\) is placed at \(\mathbf{r}_\infty=r_\infty\mathbf{n}\) in the far zone of the dielectric object, with $\mathbf{n} \cdot \mathbf{u}_\infty = 0$, and producing fields \((\mathbf{E}_\infty,\mathbf{H}_\infty)\). The origin of the reference system is in the centroid of the dielectric object, that is, the center of the sphere circumscribing the dielectric object with minimum radius $\ell_c$.}
    \label{fig:Lorentz}
\end{figure}

\subsection{Lorentz Reciprocity and a far-field identity}

\label{sec:Lorentz}

We derive here an identity used later in Sec.~\ref{sec:RadiatedPower}. 
Consider two source–field configurations at the same
frequency $\omega$ in the presence of the dielectric object, as depicted in Fig. \ref{fig:Lorentz}. We choose as the origin of the reference system the centroid of the dielectric object, that is, the center of the sphere that circumscribes the dielectric object with minimum radius $\ell_c$.  In Scenario~I, only a dipole \(\mathbf{p}_i=\mu_i\mathbf{u}_i\) is present at \(\mathbf{r}_i\) and produces the electromagnetic field \(\mathbf{E}_i(\rb)\). In Scenario~II, a reference dipole \(\mathbf{p}_\infty=\mu_\infty\mathbf{u}_\infty\) produces the electric field \(\mathbf{E}_\infty(\rb)\). It is placed at \(\mathbf{r}_\infty=r_\infty\mathbf{n}\) in the \textit{Fraunhofer} zone (far zone) of the dielectric object, i.e. $r_\infty \gg k_\omega \ell_c^2 / \pi$,  where $\mathbf{u}_\infty$ and $\mathbf{n}$ are unit vectors with $\mathbf{n} \cdot \mathbf{u}_\infty = 0$; the associated current density is \(\mathbf{J}_\infty=-i\omega\,\mu_\infty\,\mathbf{u}_\infty\,\delta(\mathbf{r}-\mathbf{r}_\infty)\). The Lorentz reciprocity theorem yields \cite{bladel_electromagnetic_2007}
\begin{equation}
\int_{\mathbb{R}^3}\mathbf{E}_i(\rb)\cdot\mathbf{J}_\infty(\rb)\,dV
=\int_{\mathbb{R}^3}\mathbf{E}_\infty(\rb)\cdot\mathbf{J}_i(\rb)\,dV,
\end{equation}
which implies
\begin{equation}
\mu_\infty\,\mathbf{E}_i(\mathbf{r}_\infty)\cdot\mathbf{u}_\infty
=\mu_i\,\mathbf{E}_\infty(\mathbf{r}_i)\cdot\mathbf{u}_i.
\label{eq:recip-sampled}
\end{equation}
At position \(\mathbf{r}_\infty\) in the Fraunhofer zone of the dielectric objects, we have in Scenario I
\begin{equation}
\mathbf{E}_i(\mathbf{r}_\infty){\approx} \frac{\mathbf{A}_i(\mathbf{n})}{r_\infty}\,e^{ik_\omega r_\infty},
\label{eq:EfarI0}
\end{equation}
where \(\mathbf{A}_i(\mathbf{n})\) is the far-field pattern of $\mathbf{E}_i$ with $\mathbf{n}\cdot\mathbf{A}_i(\mathbf{n})=0$.
In Scenario II, the {\it incident} electric field generated at the point $\rb_i$, generated by the remote dipole alone, is
\begin{multline}
\mathbf{E}_\infty^{inc}(\rb_i)
\approx \frac{k_\omega^2}{4\pi\varepsilon_0}\,\frac{\mu_\infty}{r_\infty} \exp\left(ik_\omega\,\mathbf{n}\cdot \mathbf{r}_i\right) \exp\left(-ik_\omega r_\infty\right) \mathbf{u}_\infty, 
\end{multline}
as $\rb_i$ lies in the Fraunhofer zone of the electromagnetic field generated by the dipole $\mathbf{p}_\infty$. Thus, in Scenario II, the dielectric object experiences an incident field that is a transverse plane wave propagating along $\mathbf{n}$, polarized along $\mathbf{u}_\infty$ and with unit amplitude provided that
$\mu_\infty=4\pi\varepsilon_0\,r_\infty/{k_\omega^2}$. According to Eqs. \ref{eq:Ffield}-\ref{eq:FfieldBC}, we denote by $\mathbf{F}_{\omega\mathbf{n}\mathbf{u}_\infty}(\mathbf{r}_i)$ the \textit{total} electric field at $\rb_i$ scattered by the dielectric objects when excited by a plane wave of unit intensity propagating along $\mathbf{n}$ and polarized along $\mathbf{u}_\infty$, hence 
\begin{equation}
    \mathbf{E}_\infty (\mathbf{r}_i) = \mathbf{F}_{\omega\mathbf{n}\mathbf{u}_\infty}(\mathbf{r}_i)
    \label{eq:RelazioneEF}.
\end{equation}
This equation is valid up to the overall phase factor $\exp\left({-ik_\omega r_\infty}\right)$ that cancels in the power observables.
Using this result in \eqref{eq:recip-sampled} together with \eqref{eq:EfarI0} we obtain the projection identity
\begin{equation}
\mathbf{A}_i(\mathbf{n})\cdot\mathbf{u}_\infty
=\frac{k_\omega^2}{4\pi\varepsilon_0}\,\mu_i\,
\mathbf{F}_{\omega\mathbf{n}\mathbf{u}_\infty}(\mathbf{r}_i)\cdot\mathbf{u}_i,
\label{eq:relationFA0}
\end{equation}
 In the scattering-mode normalization introduced in Eq.~\eqref{eq:ScatteringModes}, Eq. \ref{eq:relationFA0} becomes:
\begin{equation}
\mathbf{A}_i(\mathbf{n})\cdot\mathbf{u}_\infty
=\sqrt{\frac{\pi\omega}{\hbar\varepsilon_0 c}}\,
\mu_i\,[\mathbf{E}_{\omega\mathbf{n}\mathbf{u}_\infty}(\mathbf{r}_i)\cdot\mathbf{u}_i].
\label{eq:relationFA-modes0}
\end{equation}
This relation states that the \(\mathbf{u}_\infty\)-polarized far-field amplitude in direction \(\mathbf n\) generated by the dipole at $\mathbf{p}_i$ is proportional to the projection of the normalized scattering mode \(\mathbf{E}_{\omega\mathbf n\mathbf{u}_\infty}\) at the dipole location onto the dipole orientation \(\mathbf u_i\). In other words, it links the coupling between the dipole and the scattering mode at \(\mathbf r_i\) to the corresponding far-field pattern produced when the object is driven by a dipole \(\mu_i\mathbf u_i\) at \(\mathbf r_i\).

\subsection{Time-averaged power emitted by the dipoles}

The time-averaged power emitted by the $N$ dipoles is

\begin{equation}
    \Power^{em} = - \frac{1}{2} \text{Re} \int_{\mathbb{R}^3} d^3 \mathbf{r} \, \mathbf{J}^* \cdot \mathbf{E} = \sum_{i,j=1}^N \Power_{ij}^{em},
\end{equation}
where
\begin{equation}
    \Power^{em}_{ij} = - \frac{1}{2} \text{Re} \int_{\mathbb{R}^3} d^3 \mathbf{r} \, \mathbf{J}_i^*\cdot \mathbf{E}_j.
    \label{eq:Pemij}
\end{equation}
Using \ref{eq:Etot} in \ref{eq:Pemij}, one finds the following
\begin{equation}
    \Power_{ij}^{em}  = \frac{\mu_0}{2} \omega^3  (\mu_i \mathbf{u}_i) \cdot \text{Im} \left[ \mathcal{G}_\omega (\mathbf{r}_i, \mathbf{r}_j) \right] \cdot (\mu_j \mathbf{u}_j).
\end{equation}
A comparison of the above equation with Eq.~\eqref{eq:gequiv1} yields
\begin{equation}
    \Power_{ij}^{em}  = \frac{\hbar\omega}{4}  \Gamma_{ij}.
\end{equation}

\subsection{Time-averaged power absorbed by the dielectric object}
The time-averaged power absorbed by the dielectric objects is

\begin{multline}
    \Power^{abs} = \frac{\omega}{2} \varepsilon_0  \int_{V} d^3 \mathbf{r} \, \text{Im} [\varepsilon_\omega (\mathbf{r}) ] \mathbf{E}(\mathbf{r}) \cdot \mathbf{E}^* (\mathbf{r}) \\  = \sum_{i,j=1}^N \Power^{abs}_{ij},
    \label{eq:PowerAbs}
\end{multline}
with
\begin{equation}
    \Power^{abs}_{ij} = \frac{\omega}{2} \varepsilon_0  \int_{V} d^3 \mathbf{r} \, \text{Im} [\varepsilon_\omega (\mathbf{r}) ] \mathbf{E}_i (\mathbf{r}) \cdot \mathbf{E}^*_j (\mathbf{r}).
    \label{eq:PowerAbsij}
\end{equation}
Using \ref{eq:Etot}  and the reciprocity for the dyadic Green function, we get:

\begin{multline}
    \Power^{abs}_{ij} = \frac{1}{2} \varepsilon_0 \mu_0^2 \omega^5  \mu_i \mu_j  \times \\ \int_{V} d^3 \text{Im} [\varepsilon_\omega (\mathbf{r}) ]  \mathbf{u}_i \cdot  \mathcal{G}_\omega (\mathbf{r}_i,\mathbf{r}) \mathcal{G}_\omega^{*T} (\mathbf{r}_j,\mathbf{r})  \cdot \mathbf{u}_j,
\end{multline}
where we have used the property $\mathcal{G}_\omega(\mathbf{r},\mathbf{r}_j) = \mathcal{G}_\omega^T(\mathbf{r}_j,\mathbf{r})$. Using the relation between $\mathcal{G}_e$ and  $\mathcal{G}_\omega$ in Eq. \ref{eq:Ge}, we get: 

\begin{multline}
    \Power^{abs}_{ij} = \frac{\pi }{2} {\hbar \omega } \frac{\mu_i \mu_j}{\hbar^2} \times \\ \int_{V} d^3 \mathbf{r} \, \mathbf{u}_i \cdot  \mathcal{G}_e (\mathbf{r}_i,\mathbf{r}, \omega) \mathcal{G}_e^{*T} (\mathbf{r}_j,\mathbf{r},\omega)  \cdot \mathbf{u}_j = \frac{\pi}{2} {\hbar \omega} \mathcal{J}^{(M)}_{ij},
    \label{eq:PabsJ}
\end{multline}
where \(\mathcal{J}^{(M)}_{ij}\) is defined in Eq.~\eqref{eq:Jmat}. 

\subsection{Averaged Power Radiated to infinity}
\label{sec:RadiatedPower}

The time-averaged power radiated through a sphere \(S_\infty\) at infinity is

\begin{multline}
\Power^{rad} = \frac{1}{4} \int_{S_\infty} \, do_{\mathbf{n}}  r_\infty^2  \left( \mathbf{E} \times \mathbf{H}^* + \mathbf{E}^* \times \mathbf{H} \right) \cdot \mathbf{n}  = \\
\sum_{i,j=1}^N \Power^{rad}_{ij},
\label{eq:Prad}
\end{multline}
with
\begin{equation}
    \Power^{rad}_{ij} =  \oint_{S_\infty} \, do_{\mathbf{n}}  r_\infty^2 \mathbf{S}_{ij} (r_\infty, \mathbf{n}) \cdot \mathbf{n},
    \label{eq:Pradij}
\end{equation}
and the \textit{symmetrized mutual Poynting vector} is defined as:
\begin{equation}
    \mathbf{S}_{ij} = \frac{1}{4} \left( \mathbf{E}_i \times \mathbf{H}_j^* + \mathbf{E}_j^* \times \mathbf{H}_i \right).
    \label{eq:mutualPoynting}
\end{equation}
In the far zone, we have
\begin{multline}
\mathbf{H}_{i} (\mathbf{r}_\infty)= \frac{1}{\zeta_0} \hat{\mathbf{n}} \times \mathbf{E}_{i} ({r}_\infty,\mathbf{n}) \\ = \frac{1}{\zeta_0}  \hat{\mathbf{n}} \times \mathbf{A}_i (\mathbf{n}) \, \frac{\exp{ \left( i k_\omega {r}_\infty \right) }}{r_\infty}.
\label{eq:Hfar}
\end{multline}
The Poynting vector \ref{eq:mutualPoynting} becomes
\begin{equation}
    \mathbf{S}_{ij} = 
    \frac{1}{2 \zeta_0} \frac{1}{r_\infty^2} \left[  \mathbf{A}_i (\mathbf{n}) \cdot   \mathbf{A}_j^* (\mathbf{n})  \right] \, \mathbf{n}.
\end{equation}
By replacing this expression in \ref{eq:Pradij} we get:
\begin{equation}
\label{eq:PradJ}
\Power^{rad}_{ij} = \frac{1}{2\zeta_0}  \oint_{S_\infty} \, do_{\mathbf{n}} \, \displaystyle \sum_{\nu} ( \mathbf{A}_i \cdot \boldsymbol{\nu} ) ( \mathbf{A}_j^* \cdot \boldsymbol{\nu}).
\end{equation}
Using the relation \ref{eq:relationFA-modes0} one finds
\begin{multline}
\Power^{rad}_{ij} = \frac{\pi}{2}
{\hbar  \omega} \frac{ \mu_i  \mu_j}{\hbar^2} \times \\  \oint_{S_\infty} \, do_{\mathbf{n}} \, \sum_\nu \mathbf{u}_i \cdot \mathbf{E}_{\omega \mathbf{n} \nu} (\mathbf{r}_i)  \mathbf{E}^*_{\omega \mathbf{n} \nu} (\mathbf{r}_j) \cdot \mathbf{u}_j =  \frac{\pi}{2}
{\hbar  \omega} \mathcal{J}_{i j}^{(S)}(\omega),
\end{multline}
where \(\mathcal{J}^{(S)}_{ij}\) is defined in Eq.~\eqref{eq:Jsca}.

In summary, the spectral-density matrices \(\mathcal J^{(S)}_{ij}\) and \(\mathcal J^{(M)}_{ij}\) can be computed purely within classical electrodynamics by evaluating the pairwise radiated and absorbed powers \(\Power^{rad}_{ij}\) and \(\Power^{abs}_{ij}\).

\subsection{Mutual Poynting theorem}

Direct evaluation of Eq.~\eqref{eq:PabsJ} requires a volume integral over \(V\), and hence a volumetric mesh. This can be avoided by recasting the calculation as a surface integral. 

\begin{figure}
    \centering
    \includegraphics[width=0.5\linewidth]{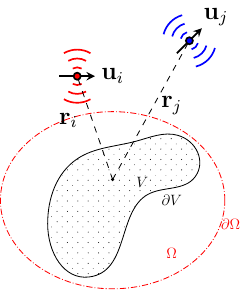}
    \caption{Proof of the mutual Poynting theorem. \(\Omega\) is any volume whose boundary \(\partial\Omega\) encloses the dielectric body but excludes the sources \(i\) and \(j\). }
    \label{fig:Poynting}
\end{figure}

Let \(\Omega\) be any volume whose boundary \(\partial\Omega\) encloses the dielectric body but excludes the sources \(i\) and \(j\) (Fig.~\ref{fig:Poynting}). Using Maxwell’s equations in \(\Omega\) and taking the divergence of \(\mathbf S_{ij}\) in \eqref{eq:mutualPoynting} yields the mutual Poynting theorem

\begin{multline}
   \mathscr{P}_{ij}^{abs} = \frac{\omega}{2} \varepsilon_0 \int_V d^3 \mathbf{r} \, \text{Im} [ \varepsilon_\omega (\mathbf{r}) ] \mathbf{E}_i(\mathbf{r}) \cdot \mathbf{E}_j^*(\mathbf{r}) \\ =
    - \oint_{\partial \Omega} d^2 \mathbf{r} \, \mathbf{S}_{ij} \, \cdot \mathbf{m}, 
\label{eq:mutualPoyntingTheorem}
\end{multline}
where $\mathbf{m}$ denotes the outward-pointing unit normal vector on $\partial\Omega$. Equation~\eqref{eq:mutualPoyntingTheorem} allows one to rewrite the volume integral entering \(\mathcal{J}^{(M)}_{ij}\) as a surface integral over \(\Sigma\), which is advantageous for numerical evaluation with a surface-integral-equation solver.

\section{Correlators}
\label{sec:Correlators}
In this Section we compute the elements $C_{ij}^{(M)}(\tau)$ of the correlator matrix of the medium-assisted reservoir; a similar result holds for the elements of the correlator matrix of the scattering-assisted reservoir.

We consider the following scenario: (a) the initial quantum state of the electromagnetic environment is a product state, that is, $\hat{\rho}_E(0)=\hat{\rho}_E^{(M)}(0) \otimes \hat{\rho}_E^{(S)}(0)$, where $\hat{\rho}_E^{(M)}(0)$ and $\hat{\rho}_E^{(S)}(0)$ are the initial density operators of the medium-assisted reservoir and the scattering-assisted reservoir, respectively; (b) the two reservoirs are initially in thermal states at temperatures $T_0^{(M)}$ and $T_0^{(S)}$.

The correlation function $C^{(M)}_{ij}(\tau)$ is given by
\begin{equation}
\label{eq:corr}
    C^{(M)}_{ij}(\tau)=\langle\hat{F}^{(M)}_i(\tau) \hat{F}^{(M)}_j(0)\rangle
\end{equation}
where $\hat{{F}}^{(M)}_i(\tau)$ is the interaction operator of the medium-assisted reservoir in the interaction picture,
\begin{equation}
\hat{{F}}^{(M)}_i(\tau) = \hat{U}^\dagger_E(\tau)\hat{F}^{(M)}_i \hat{U}_E(\tau),
\label{eq:EvolutionF}
\end{equation}
$\hat{U}_E(\tau)=\exp(-i\hat{H}_{E}^\text{bright}\tau/\hbar)$ is the free evolution operator of the bright bosonic modes of the electromagnetic environment, and $\langle \cdot \rangle = \text{Tr}_{E_M}[\cdot\,\rho_E^{(M)}(0)]$. Starting with \ref{eq:EvolutionF}, and using Eqs. \ref{eq:F_M}, \ref{eq:HEbright}, and the commutation relations \ref{eq:commut_C} we obtain
\begin{equation}
\label{eq:F_M1}
\hat{F}_i^{(M)}(\tau)=\sum_{j=1}^N \int_0^\infty d\omega [G_{ij}^{(M)}(\omega)\hat{C}_{j}(\omega) e^{-i\omega \tau}+h.c.].
\end{equation}
Substituting \ref{eq:F_M1} into \ref{eq:corr}, and using again \ref{eq:commut_C} we find
\begin{widetext}
\begin{equation}
    C_{ij}^{(M)}(\tau)=\int_0^\infty d\omega \sum_{n=1}^N \left[G^{(M)}_{in}(\omega)G^{(M)*}_{jn}(\omega)\langle \hat{C}_n(\omega)\hat{C}^\dagger_n(\omega)\rangle e^{-i\omega \tau} +
    G^{(M)*}_{in}(\omega)G^{(M)}_{jn}(\omega)\langle \hat{C}^\dagger_n(\omega)\hat{C}_n(\omega)\rangle e^{+i\omega \tau} \right],
    \label{eq:CijM}
\end{equation}
\end{widetext}
because for a thermal state
$
\langle \hat C_m(\omega)\hat C_n(\omega)\rangle 
=\langle \hat C_m^{\dagger}(\omega)\hat C_n^{\dagger}(\omega)\rangle =0,
$
$ \langle \hat C_m(\omega)\hat C_n^{\dagger}(\omega)\rangle 
=\delta_{mn}\big(1+n_\omega^{(M)}\big)$, $
\langle \hat C_m^{\dagger}(\omega)\hat C_n(\omega)\rangle
=\delta_{mn}\,n_\omega^{(M)},$ where 
\begin{equation}
n_\omega^{(M)}=\langle \hat{C}_n^\dagger(\omega)\hat{C}_n(\omega)\rangle=\frac{1}{e^{\beta_{M}\hbar\omega}-1},
\end{equation}
and $\beta_{M}=1/(k_BT_0^{(M)})$.
Exploiting that
\begin{equation}
\langle \hat{C}_n(\omega)\hat{C}_n^\dagger(\omega)\rangle=1+\langle \hat{C}_n^\dagger(\omega)\hat{C}_n(\omega)\rangle,
\end{equation}
$G_{jn}^{(M)*}(\omega)=G_{nj}^{(M)}(\omega)$ and $\sum^N_{ n=1} G^{(M)}_{in}(\omega)G^{(M)}_{nj}(\omega)=M_{ij}$, Eq. \ref{eq:CijM} becomes
\begin{multline}
    C_{ij}^{(M)}(\tau)= \\ \int_0^{\infty} \mathrm{d} \omega [(1+n_\omega^{(M)}) {M}_{ij}^{}(\omega)e^{-i\omega \tau}+
    n_\omega ^{(M)}{M}_{ij}^{*}(\omega)e^{i\omega \tau}].
\end{multline}

\section{Surface Integral Equation Method}
\label{sec:SIE}
Surface–integral–equation (SIE) formulations \cite{harrington_field_1993} are well suited for the classical evaluation of spectral-density matrices when the dielectric object is piecewise homogeneous: unknowns live only on object boundaries and the radiation condition at infinity is inherently satisfied. 

We consider a single dielectric body that occupies a bounded volume \(V\) with a boundary \(\partial V\). The interior (``\(+\)'') medium is homogeneous, with \(\varepsilon^+(\omega)=\varepsilon_0\,\varepsilon_\omega\) and \(\mu^+=\mu_0\). The exterior (``\(-\)'') medium is vacuum, with \(\varepsilon^-=\varepsilon_0\) and \(\mu^-=\mu_0\). Let \(k_\omega^\pm=\omega\sqrt{\mu^\pm\varepsilon^\pm}\) and \(\zeta^\pm=\sqrt{\mu^\pm/\varepsilon^\pm}\) denote, respectively, the wavenumbers and wave impedances. The object is illuminated by a time–harmonic field, \(\mathrm{Re}\{\mathbf E_{\mathrm{inc}}(\mathbf r)\,e^{-i\omega t}\}\), \(\mathrm{Re}\{\mathbf H_{\mathrm{inc}}(\mathbf r)\,e^{-i\omega t}\}\).

Following Poggio–Miller–Chang–Harrington–Wu–Tsai (PMCHWT) \cite{wu_scattering_1977,chang_surface_1977,poggio_chapter_1973,forestiere_surface_2012}, the equivalent electric and magnetic surface currents \(\mathbf j_e(\mathbf r)\) and \(\mathbf j_m(\mathbf r)\), defined on \(\partial V\), solve:

\begin{equation}
 \boldsymbol{\mathcal{Z}} \,  \mathbf{j} = \mathbf{v},
  \label{eq:PMCHWT}
\end{equation}
with block operators
\begin{equation}
  \boldsymbol{\mathcal{Z}}  = \left( {\begin{array}{*{20}c}
    \Ze \Te +  \Zi \Ti &  \Ke +\Ki
  \\
      -  \left( \Ke +\Ki \right) &  
\Te/\Ze  + \Ti/\Zi  \\
\end{array}} \right),
  \label{eq:PMCHWTbis}
\end{equation}
$ \mathbf{j} =  \left[ \je,  \jm \right]^T$,
$\mathbf{v} = \left[     \eo,  \ho  \right]^T$,
$ \mathbf{e}_0  =  \left. - \mathbf{n} \times \mathbf{n} \times {\bf{E}}_{inc}  \right|_{\partial V}$, and $
     \mathbf{h}_0  =  \left. - \mathbf{n} \times \mathbf{n} \times {\bf{H}}_{inc}  \right|_{\partial V}.$
The EFIE/MFIE boundary operators \(\mathcal T_\pm\) and \(\mathcal K_\pm\) act on a tangential test function \(\mathbf w\) as: 
\begin{subequations}
  \begin{align}
    \label{eq:Operator_K}
    \Kie & \left\{ {\bf w} \right\} \left( {\bf r} \right) = \mathbf{n}\times \mathbf{n}\times \int_{\partial V} {\bf w}  \left( {\bf r}' \right) \times \nabla' g^\pm \left( {\bf
r}  - {\bf r}' \right) dS',  \\
    \label{eq:Operator_T}
    \notag
    \Tie & \left\{ {\bf w} \right\} \left( {\bf r} \right) =    -i k^\pm_\omega \mathbf{n}\times \mathbf{n}\times  \int_{\partial V} g^\pm \left( \dr \right) {\bf w} \left( {\bf r}' \right) dS'\\ & - \frac{1}{i k^\pm_\omega} \mathbf{n}\times \mathbf{n}\times  \int_{\partial V} \nabla' g^\pm \left( \dr \right) \nabla' _S  \cdot {\bf w}\left( {\bf r}' \right) dS',
    \end{align}
\end{subequations}
where
\begin{equation}
  g^\pm \left( {\bf r} - {\bf r}' \right) = \frac{ e^{i k^\pm_\omega \left|{\bf r} - {\bf r}' \right|} }{4 \pi \left|{\bf r} - {\bf r}' \right| }.
  \label{eq:GreenFunctin}
\end{equation}
is the homogeneous-space Green function.

\subsection{Finite-dimensional representation}
\label{sec:Representation}
\begin{figure}
    \centering
    \includegraphics[width=0.75\linewidth]{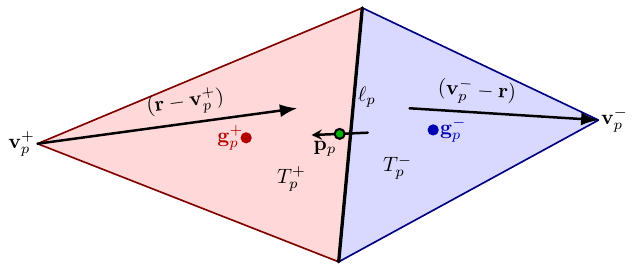}
    \caption{Illustration of the RWG basis function associated to the $p$ th edge and defined on the triangle pair $T_p^{+}, T_p^{-}$ with centroids \(\mathbf g_p^{\pm}\). In $T_p^{+}$and $T_p^{-}$ the vector field $\mathbf{f}_p$ is proportional to the vector ( $\mathbf{r}-\mathbf{v}_p^{+}$) and ( $\mathbf{v}_p^{-}-\mathbf{r}$ ), respectively. To each RWG basis function is associated a dipole moment $\mathbf{p}_p$ centered in the midpoint \(\mathbf c_p=(\mathbf g_p^{+}+\mathbf g_p^{-})/2\). }
    \label{fig:RWG}
\end{figure}

We discretize \(\partial V\) with a conforming triangular mesh \(\mathcal M\) having \(N_e\) interior edges. For edge \(p\), let \(\ell_p\) be its length; \(T_p^{\pm}\) the adjacent triangles with areas \(A_p^{\pm}\); \(\mathbf v_p^{\pm}\) the vertices opposite to the common edge; and \(\mathbf g_p^{\pm}\) their centroids. Define the centroid offset \(\mathbf d_p=\mathbf g_p^{+}-\mathbf g_p^{-}\) and the midpoint \(\mathbf c_p=(\mathbf g_p^{+}+\mathbf g_p^{-})/2\). The Rao–Wilton–Glisson (RWG) basis function \(\mathbf f_p\) is the piecewise linear tangential field
$$
\mathbf{f}_p(\mathbf{r})= \frac{\ell_p}{2} \times
\begin{cases}
\left(\mathbf{r}-\mathbf{v}_p^{+}\right)/A_p^{+} & \mathbf{r} \in T_p^{+} \\
\left(\mathbf{v}_p^{-}-\mathbf{r} \right)/A_p^{-} & \mathbf{r} \in T_p^{-} \\
0 & \text{elsewhere}
\end{cases} 
$$
whose support is $S_p = T_p^{+} \cup T_p^{-}$. Its surface divergence is constant over each triangle. Therefore, by the continuity equation, the RWG basis function $\mathbf{f}_p$ is associated with the total electric charge $Q_p^{\pm}=\pm \ell_p/(i\omega)$ on $T_p^{\pm}$, which can be regarded as localized at the corresponding centroid $\mathbf{g}_p^{\pm}$, and with the total magnetic charge $Q_m^{\pm}=\pm \ell_p/(i\omega\mu_0)$, likewise localized at $\mathbf{g}_p^{\pm}$.

We expand the unknown currents in the RWG basis functions \(\{\mathbf f_p\}_{p=1}^{N_e}\):
\begin{equation}
\mathbf{j}_e (\mathbf{r}) \approx \sum_{p=1}^{N_e} \alpha_p \mathbf{f}_p(\mathbf{r}), \qquad
\mathbf{j}_m (\mathbf{r}) \approx \sum_{p=1}^{N_e} \beta_p \mathbf{f}_p(\mathbf{r}).
\end{equation}
Galerkin testing with the same basis yields the finite–dimensional counterpart of \eqref{eq:PMCHWT}:
\begin{equation}
\mathbf{Z} \, \mathbf{J} = \mathbf{V}
\label{eq:PMCHWT_finite}
\end{equation}
with
\begin{equation}
	\mathbf{Z} = \begin{bmatrix}
		\zeta^- T_- + \zeta^+ T_+   &K_- + K_+\\
		-(K_- + K_+) & \frac{1}{\zeta^-} T_- + \frac{1}{\zeta^+} T_+
	\end{bmatrix},
 \label{eq:Zpq_delta}
\end{equation}
\begin{equation}
[T_\pm]_{ij} = \langle \mathbf{f}_i | \mathcal{T}_\pm | \mathbf{f}_j \rangle, \hspace{5ex} 	[K_\pm]_{ij} = \langle \mathbf{f}_i | \mathcal{K}_\pm | \mathbf{f}_j \rangle,
 \label{eq:PMCHWT_Finite1}
\end{equation}
$ \vec{J}=  [ \vec{J}_e, \vec{J}_m ]^T$, $ \vec V = \left[ \vec{E}_0, \vec{H}_0 \right]^T
$,
$\vec{J}_{e} = [\alpha_1, \ldots ,\alpha_{N_e}]^{\intercal}$,  
$\vec{J}_{m} = [\beta_1, \ldots ,\beta_{N_e}]^{\intercal}$, $[\vec{E}_0]_i = \langle \mathbf{f}_i | \vec{e}_0 \rangle$, 
$[\vec{H}_0]_i = \langle \mathbf{f}_i | \vec{h}_0 \rangle$, and
\begin{equation}
	\langle \vec{u} | \vec{v} \rangle = \int_{\partial V} d^2 \mathbf{r}   \vec{u}^*(\vec{r}) \cdot  \vec{v}(\vec{r}) .
\end{equation}
To evaluate pairwise absorbed and radiated powers \(\Power^{\mathrm{abs}}_{ij}\) and \(\Power^{\mathrm{rad}}_{ij}\), many right–hand sides are required (one per impressed dipole). When \(N\) is large, it is efficient to compute a single LU factorization of \(\mathbf Z\)  and reuse it for all right–hand sides by matrix-vector multiplications. We denote $\{\alpha_p^{(i)}\}_{p\in N_e}$ $\{\beta_p^{(i)}\}_{p\in N_e}$ the expansion coefficients for the $i$-th impressed dipole excitation.
 
\subsection{Time-averaged power absorbed by the dielectric object}

We evaluate \(\mathscr{P}^{\mathrm{abs}}_{ij}\) via the mutual Poynting theorem, Eq.~\eqref{eq:mutualPoyntingTheorem}. A natural choice for \(\partial\Omega\) is the particle boundary \(\partial V\), where the fields can be obtained directly from the solved equivalent surface currents (and the associated surface charges). However, for irregularly shaped bodies the surface fields on \(\partial V\) may be large and rapidly varying, which can slow the convergence of the surface integral. In such cases it is advantageous to choose a smooth enclosing surface \(\partial\Omega\) , e.g. a spherical/spheroidal surface, that surrounds \(V\) but lies a distance away from it, provided \(\partial\Omega\) encloses \(V\) and excludes the impressed sources.

\subsection{Average Powers radiated to infinity}
To compute the $\mathscr{P}_{ij}^{rad}$
we solve the SIE scattering problem for two excitations: an impressed dipole \(\mathbf p_i\) at \(\mathbf r_i\) and, separately, \(\mathbf p_j\) at \(\mathbf r_j\). Denote the corresponding RWG expansion coefficients by \(\{\alpha_s^{(i)},\beta_s^{(i)}\}\) and \(\{\alpha_t^{(j)},\beta_t^{(j)}\}\).
Then, a brute-force but expensive procedure would evaluate the far-field surface integral in
Eq.~\eqref{eq:PradJ} by forming the radiation from the equivalent surface
currents at all quadrature points on \(S_\infty\) for both \(\mathbf p_i\) and \(\mathbf p_j\) excitations, which is an
expensive procedure.

Instead, \(\mathscr{P}_{ij}^{\mathrm{rad}}\) can be obtained analytically from the RWG
expansion coefficients. As noted in Sec.~\ref{sec:Representation}, the RWG basis function $\mathbf{f}_p$ induces equal and opposite electric or magnetic charges on \(T_p^\pm\), thus a single coefficient \(\alpha_p\) or \(\beta_p\)  corresponds to equivalent electric or magnetic dipole moments (centered at \(\mathbf c_p\), oriented along \(\mathbf d_p\))
\begin{equation}
 \mathbf{p}_p^{(i)} = \alpha_p^{(i)} \frac{\ell_p}{i \omega} \mathbf{d}_p, \quad 
 \mathbf{m}_p^{(i)} =  \beta_p^{(i)} \frac{  \ell_p}{ i \omega \mu_0} \mathbf{d}_p, 
\end{equation}
with $p=1,\ldots,N_e$. For easy of reference, set  $\mathbf{p}_0^{(i)} = \mathbf{p}_i$ and $\mathbf{c}_0^{(i)} = \mathbf{r}_i$ (There is no impressed magnetic dipole).

The mutual radiated power is then the sum of free–space pairwise contributions between the effective dipoles of the two solutions
\begin{multline}
    \mathscr{P}_{ij}^{rad} = \displaystyle\sum_{s,t=0}^{N_e}  \mathcal{P}_{ee} (\mathbf{p}_s^{(i)},\mathbf{p}_t^{(j)}) + \displaystyle\sum_{s,t=1}^{N_e} \mathcal{P}_{mm}(\mathbf{m}_s^{(i)},\mathbf{m}_t^{(j)})  
    \\ + \displaystyle\sum_{s,t=1}^{N_e} \left[ \mathcal{P}_{em}(\mathbf{p}_s^{(i)},\mathbf{m}_t^{(j)}) + \mathcal{P}_{em}(\mathbf{m}_s^{(i)},\mathbf{p}_t^{(j)})\right].
\end{multline}
The first sum takes into account the pairwise
interactions of the electric dipoles, including the impressed ones that occur for $s=0$ and $t=0$:
\begin{multline}
\mathcal{P}_{ee}(\mathbf{p}_s,\mathbf{p}_t)  = \frac{k_\omega^{4}}{8\pi \varepsilon_{0}^2 \zeta_0} \times \\ \left\{  \mathbf{p}_s \cdot \mathbf{p}_t^* f(k_\omega r_{st})  -  \left( \mathbf{p}_s \cdot  \hat{\mathbf{r}}_{st}\right) \left( \mathbf{p}_t^* \cdot  \hat{ \mathbf{r}}_{st} \right) g (k_\omega r_{st})    \right\}.
\end{multline}
The second sum takes into account the pairwise
interactions of magnetic dipoles, where $\mathcal{P}_{mm}$ is directly related to $\mathcal{P}_{ee}$ by the duality property of the electromagnetic field:
\begin{equation}
    \mathcal{P}_{mm}^{ij} (\mathbf{m}_s,\mathbf{m}_t)  = \frac{1}{c^2} \mathcal{P}_{ee}^{ij} (\mathbf{m}_s,\mathbf{m}_t).
\end{equation}
The third sum takes into account the pairwise 
interactions of the electric and magnetic dipoles and viceversa with
\begin{equation}
\mathcal{P}_{em}^{ij} (\mathbf{p}_s,\mathbf{m}_t)  = i \frac{1}{c} \frac{k_\omega^{4}}{8\pi \varepsilon_{0}^2 \zeta_0} \mathbf{p}_s \times \mathbf{m}^*_t \cdot \hat{\mathbf{r}}_{st} \, h(k_\omega r_{st}) 
\end{equation}
where ${\mathbf{r}}_{st} = \mathbf{c}_s -\mathbf{c}_t $, ${r}_{st} = \left| \mathbf{r}_{st} \right|$, 
$\hat{\mathbf{r}}_{st} = {\mathbf{r}}_{st}/r_{st}$, and
\begin{equation}
\left\{
\begin{aligned}
f(x) &=  \frac{\sin x}{x} - \, \frac{ \sin x - x \cos x }{x^3}, \\
g(x)&=\frac{\sin x}{x} - 3 \, \frac{ \sin x - x \cos x }{x^3}, \\
h(x)&=\frac{\sin(x) - x \cos (x)}{x^2},
\end{aligned}
\right.
\end{equation}
$ f(x\rightarrow 0) = {2}/{3}$, $ g(x\rightarrow 0) = 0$, and  $ h(x\rightarrow 0) = 0$.

\section{Lorentz Reciprocity and a far-field identity (alternative proof)}

We derive here an identity used in Sec.~\ref{sec:RadiatedPower}.
Consider two source–field configurations at the same
frequency $\omega$ in the presence of the dielectric object, as depicted in Fig. \ref{fig:Lorentz}.  In \textit{Scenario}~I, only a dipole \(\mathbf{p}_i=\mu_i\mathbf{u}_i\) is present at \(\mathbf{r}_i\) and produces the electromagnetic field \((\mathbf{E}_i,\mathbf{H}_i)\).  At position \(\mathbf{r}_\infty\) at infinity  
\begin{subequations}
\begin{align}
\mathbf{E}_i(\mathbf{r}_\infty) &= \frac{e^{ik_\omega r_\infty}}{r_\infty}  \mathbf{A}_i(\mathbf{n})  + \mathcal{O}(r_\infty^{-2}),  \, \\
\mathbf{H}_i(\mathbf{r}_\infty) &=  \frac{e^{ik_\omega r_\infty}}{r_\infty} \frac{\mathbf{n} \times \mathbf{A}_i(\mathbf{n})}{\zeta_0} + \mathcal{O}(r_\infty^{-2}),
\end{align}
\label{eq:EfarI}
\end{subequations}
where \(\mathbf{A}_i(\mathbf{n})\) is the far-field pattern of $\mathbf{E}_i$ with $\mathbf{n}\cdot\mathbf{A}_i(\mathbf{n})=0$.

In \textit{Scenario~II}, the object is illuminated by a unit-amplitude plane wave with propagation direction $\mathbf{n}_\infty$ and polarization $\mathbf{u}_\infty$:
\begin{subequations}
\begin{align}
    \mathbf{E}_2^{inc} &= \exp \left( i k_\omega \mathbf{n}_\infty \cdot \mathbf{r} \right) \mathbf{u}_\infty, \\
    \mathbf{H}_2^{inc} &= \frac{1}{\zeta_0} \mathbf{n}_\infty \times \mathbf{u}_\infty \exp \left( i k_\omega \mathbf{n}_\infty \cdot \mathbf{r} \right),
\end{align}
\end{subequations}
By linearity, the total fields are
\begin{equation}
    \mathbf{E}_2 = \mathbf{E}_2^{inc} + \mathbf{E}_2^{sca}, \qquad
    \mathbf{H}_2 = \mathbf{H}_2^{inc} + \mathbf{H}_2^{sca},
\end{equation}
with scattered fields $\mathbf{E}_2^{sca}$ and $\mathbf{H}_2^{sca}$ satisfying the Silver–Müller condition,
\begin{subequations}
\begin{align}
\mathbf{E}^{sca}_2 (\mathbf n, r_\infty)&=\frac{e^{i k_\omega r_\infty}}{r_\infty}\,\mathbf E_\infty({\mathbf n} )+\mathcal{O}(r_\infty^{-2}), \\
\mathbf H_2^{sca}(\mathbf n, r_\infty)&=\frac{e^{ik_\omega r_\infty}}{r_\infty}\,\frac{\mathbf{n}\times\mathbf E_\infty( {\mathbf n} ) }{\zeta_0}+\mathcal{O}(r_\infty^{-2}).
\end{align}
\label{eq:SilverMuller}
\end{subequations}
Note that the total electric field $\mathbf{E}_2$ coincides with $\mathbf{F}_{\omega\mathbf{n}_\infty\mathbf{u}_\infty}$ as defined in Eqs.~\eqref{eq:Ffield}–\eqref{eq:FfieldBC}.

The Lorentz reciprocity theorem gives
\begin{equation}
    \oint_{S_\infty} d^2  \mathbf{r} \, \left(\mathbf{E}_i \times \mathbf{H}_{2}-\mathbf{E}_{2} \times \mathbf{H}_i\right) \cdot \mathbf{n}   =\int_{\mathbb{R}^3} d^3  \; \mathbf{r} \mathbf{E}_2 \cdot \mathbf{J}_i,
    \label{eq:LorentzReciprocity}
\end{equation}
where $S_\infty$ is a spherical surface at infinity and $\mathbf{J}_i$ is the current density of the dipole in Scenario~I  $\mathbf{J}_i=-i\omega\mu_i\mathbf{u}_i\,\delta(\mathbf{r}-\mathbf{r}_i)$.
Using the far-field expansions \eqref{eq:EfarI} and the Silver–Müller condition \eqref{eq:SilverMuller}, which implies that on $S_\infty$ the scattered field in Scenario~II is $\mathcal{O}(r_\infty^{-1})$ and thus negligible compared with the unit-amplitude incident plane wave-the left-hand side of \eqref{eq:LorentzReciprocity} reduces, to leading order to

\begin{multline}
   \frac{1}{\zeta_0} \frac{e^{ik_\omega r_\infty}}{r_\infty} \, \oint_{S_\infty} d^2  \mathbf{r} \, \exp \left( i k_\omega r_\infty \mathbf{n}_\infty \cdot \mathbf{n}  \right)\\ \cdot \left[   \mathbf{A}_i(\mathbf{n}) \times \mathbf{n}_\infty \times \mathbf{u}_\infty \cdot \mathbf{n} -  \mathbf{u}_\infty  \times   \mathbf{n} \times \mathbf{A}_i(\mathbf{n}) \cdot \mathbf{n} \right]. 
\end{multline}
We now use $d^2\mathbf{r}=r_\infty^2\sin\theta\,d\theta\,d\phi$:
\begin{multline}
\frac{1}{\zeta_0}{e^{ik_\omega r_\infty}} \,  \int_{0}^{2 \pi} d \phi \int_{0}^{\pi} d \theta \, {r_\infty} \sin \theta \, \exp \left( i k_\omega r_\infty \mathbf{n}_\infty \cdot \mathbf{n}  \right) \\ \cdot \left[ \mathbf{A}_i(\mathbf{n}) \times \left( \mathbf{n}_\infty \times  \mathbf{u}_\infty \right) \cdot \mathbf{n} +  \mathbf{u}_\infty  \cdot \mathbf{A}_i(\mathbf{n}) \right]. 
\label{eq:FHSLorentz}
\end{multline}

Without loss of generality, choose $\mathbf{u}_\infty = \mathbf{x}$ and $\mathbf{n}_\infty =  \mathbf{z}$. 
With $\mathbf{n}=(\sin\theta\cos\phi,\sin\theta\sin\phi,\cos\theta)$ and $\mathbf{A}_i(\mathbf{n}) \times \left( \mathbf{n}_\infty \times  \mathbf{u}_\infty \right) \cdot \mathbf{n} = \mathbf{x} \cdot  \mathbf{A}_i \cos \theta - \mathbf{z} \cdot  \mathbf{A}_i \sin \theta \cos \phi$, Eq. \ref{eq:FHSLorentz} becomes

\begin{multline}
\frac{1}{\zeta_0} {e^{ik_\omega r_\infty}} \,  \int_{0}^{2 \pi} d \phi \int_{0}^{\pi} d \theta \, {r_\infty} \sin \theta \, \exp \left( i k_\omega r_\infty \cos \theta  \right) \\ \cdot \left[    \mathbf{A}_i(\theta, \phi) \cdot \mathbf{x} (1 + \cos \theta) -  \mathbf{A}_i  (\theta, \phi) \cdot \mathbf{z} \sin \theta \cos \phi \right].
\label{eq:LHSangular}
\end{multline}
The expression~\ref{eq:LHSangular} contains integrals of the type
$$
\int_{-1}^1 d \xi \, e^{i {k}_\omega r_\infty \xi} f(\xi), 
$$
where $\xi=\cos \theta$, which can be integrated by parts to yield
$$
\frac{e^{i k_\omega r_\infty} f(1)-e^{-i k_\omega r_\infty} f(-1)}{i k_\omega r_\infty}+\mathcal{O}\left(\frac{1}{k_\omega^2 r_\infty^2}\right),
$$
provided that $d f / d \xi$ is bounded. Keeping the leading term, eq. \ref{eq:LHSangular} gives
\begin{multline}
\frac{1}{\zeta_0}  r_\infty  {e^{ik_\omega r_\infty}}  {2 \pi}  \frac{ 2 e^{ik_\omega r_\infty}}{i k_\omega r_\infty}  \left.  \mathbf{A}_i   \cdot \mathbf{x}  \right|_{\theta=0}  \\ = -i \frac{4 \pi}{\zeta_0}  \frac{1}{k_\omega} e^{2 ik_\omega r_\infty}   \left.  \mathbf{A}_i \cdot \mathbf{x}   \right|_{\theta=0}.
\end{multline}
Plugging back this expression in Eq. \ref{eq:LorentzReciprocity} and using $1/\zeta_0=\varepsilon_0 c$ and $\omega=ck_\omega$ gives 
\begin{equation}
  \left.  \mathbf{A}_i \cdot \mathbf{x}   \right|_{\theta=0}  =\frac{k_\omega^2}{ 4 \pi \varepsilon_0}  \mu_i\,\mathbf{F}_{\omega\mathbf{z}\mathbf{x}}  (\mathbf{r}_i)\cdot\mathbf{u}_i. 
\end{equation}
Going back to the original reference system, we obtain the projection identity
\begin{equation}
\mathbf{A}_i(\mathbf{n})\cdot\mathbf{u}_\infty
=\frac{k_\omega^2}{4\pi\varepsilon_0}\,\mu_i\,
\mathbf{F}_{\omega\mathbf{n}\mathbf{u}_\infty}(\mathbf{r}_i)\cdot\mathbf{u}_i,
\label{eq:relationFA}
\end{equation}
up to the overall phase factor that cancels in the power observables. In the scattering-mode normalization introduced in Eq.~\eqref{eq:ScatteringModes}, Eq. \ref{eq:relationFA} becomes:
\begin{equation}
\mathbf{A}_i(\mathbf{n})\cdot\mathbf{u}_\infty
=\sqrt{\frac{\pi\omega}{\hbar\varepsilon_0 c}}\,
\mu_i\,[\mathbf{E}_{\omega\mathbf{n}\mathbf{u}_\infty}(\mathbf{r}_i)\cdot\mathbf{u}_i].
\label{eq:relationFA-modes}
\end{equation}
Relation ~\eqref{eq:relationFA-modes} states that the \(\mathbf{u}_\infty\)-polarized far-field amplitude in direction \(\mathbf n\) is proportional to the projection of the normalized scattering mode \(\mathbf{E}_{\omega\mathbf n\boldsymbol{\nu}}\) at the dipole location onto the dipole orientation \(\mathbf u_i\). In other words, it links the local coupling between the dipole and the scattering mode  at \(\mathbf r_i\) to the corresponding far-field pattern produced when the object is driven by a dipole \(\mu_i\mathbf u_i\) at \(\mathbf r_i\).



%

\end{document}